\documentclass[twocolumn]{revtex4}

\usepackage[english]{babel} 
\usepackage{bm}
\usepackage{graphicx}
\usepackage{amsmath}
\usepackage{amssymb}
\usepackage{color}
\usepackage{comment}

\usepackage[hidelinks]{hyperref}
	
\begin{document}
		
		\title{Electron states, bound to a texture in a N\'eel antiferromagnet} 
		
		\author{N. Davier}
		\email[]{davier@irsamc.ups-tlse.fr}
		\affiliation{Laboratoire de Physique Th\'eorique, Universit\'e de Toulouse, CNRS, UPS, France}
		
		\author{R. Ramazashvili}
		\email[]{revaz@irsamc.ups-tlse.fr}
		\affiliation{Laboratoire de Physique Th\'eorique, Universit\'e de Toulouse, CNRS, UPS, France}
		
		\date{\today}
		
		\begin{abstract} 
		We study electron states, bound to topological textures such as skyrmions and domain walls in a N\'eel antiferromagnet. In certain limits, we find the dependence of bound states on the geometry of the texture, and estimate the bound-state contribution to its energy. This contribution proves significant compared with the purely magnetic energy, and thus substantially influences the shape of the texture. It also considerably shifts the transition line between the modulated and the uniform phase in favor of the latter. 
		\end{abstract}
		
		\maketitle

			\section{Introduction}
		
		Topological textures such as skyrmions and domain walls appear in various contexts from particle physics to cold atoms and, as in the present work, to solid-state magnetism \cite{Sky_Gobel_2021}. They also hold promise for spintronics \cite{Jiang2017, Sky_Back_2020, Sky_Gobel_2021} as near-future information carriers in racetrack memory devices \cite{Sky_Gobel_2021, Racetrack_Parkin2008, Racetrack_Parkin2015, Fert2013}. 

       Both ferro- and antiferromagnets have been envisaged as materials for such applications. However, in ferromagnets, Skyrmion Hall Effect (SHE) \cite{Sky_Gobel_2021} makes a skyrmion deviate from driving current, likely leading to pinning at the racetrack edges and hence information loss. Antiferromagnetic skyrmions are a promising alternative in this regard \cite{Sky_Gobel_2021,Wang, AF_Sk_Bessarab_2019,AF_Sk_Zhang_2016,AF_Sk_Gobel_2017, AF_sk_Legrand2020, AF_Sk_Barker_2016}. Firstly, zero net magnetization improves robustness: being composed of two intertwined ferromagnetic skyrmions with mutually reversed spins, antiferromagnetic skyrmion experiences no SHE, allowing rectilinear displacement. Also, antiferromagnetic skyrmions are expected to develop higher velocities than their ferromagnetic counterparts. Note, however, that skyrmions in an antiferromagnet are yet to be experimentally observed \cite{Sky_Gobel_2021, AF_Sk_Bessarab_2019,AF_Sk_Zhang_2016,AF_Sk_Gobel_2017, AF_sk_Legrand2020, AF_Sk_Barker_2016, Bogdanov_2002}. 

        Using antiferromagnetic skyrmions in future devices requires understanding their interplay with other subsystems of a solid, most notably with band electrons. Some steps in this direction have already been made: R. Cheng and Q. Niu derived quasiclassical equations of motion for the electron coordinate, momentum and spin near a texture in a N\'eel antiferromagnet \cite{Cheng_2012}. Using a different approach, 
        we have recently derived an effective-mass electron Hamiltonian in the presence of a texture \cite{Davier_2023}.  
        The latter work also showed that, for certain locations of the electron band extrema, a texture produces a peculiar (and anomalously strong) spin-orbit coupling. For a skyrmion texture, this coupling generates skyrmion-electron bound states. 
        In the presence of dopant carriers, such bound states turn the skyrmion into a charged particle.  
        Reference \cite{Davier_2023} studied a tractable example of the Belavin-Polyakov (BP) skyrmion \cite{Belavin_Polyakov_1975, Rajaraman}, and found the bound-state energy dependence on the skyrmion radius $R$ in the regimes where the bound states are shallow. 
        
        Here, we extend the results of Ref. \cite{Davier_2023} in a number of ways: 
        (i) For the BP skyrmion, we gain quantitative insight into the range where the skyrmion-electron bound states become profound, that is of energies comparable with the gap $\Delta$ in the electron spectrum of the N\'eel state. (ii) We also explore another profile, that becomes relevant in the presence of single-ion anisotropy and Dzyaloshinskii-Moriya coupling: that of a domain-wall (DW) skyrmion. Contrary to the BP skyrmion, its DW counterpart is defined by \textit{two} lengthscales rather than one: the radius $R$ and the domain wall width $w$. We study both shallow and deep bound states of an electron and a DW skyrmion. (iii) Locally, a DW skyrmion boundary is a domain wall. Thus, to gain insight, we also study band electrons in the presence of a straight domain wall, and show that it produces electron states that are bound in the transverse direction, and itinerant along the wall. This calculation also allows us to understand the genesis of electron bound states for a DW skyrmion. 
        (iv) Last but not least, at half-filling we address the influence of the skyrmion-electron bound states on the energy balance of a domain wall. 

  The key results of our work are as follows: The skyrmion-electron bound states and their energies are sensitive to the skyrmion `shape'. Moreover, the bound-state contribution to the skyrmion energy proves dominant compared with its purely `magnetic' counterpart -- and thus plays a key role in selecting an `optimal' skyrmion profile. In particular, the BP profile with a sufficiently small skyrmion radius induces no bound states, and hence no additional energy cost, which favors smaller BP skyrmions. 
  For DW skyrmions, the bound-state contribution proves dominant, as well.  
  We show how it enhances the stability range of the uniform state, and shifts the transition to the modulated phase  \cite{Bogdanov_1989} to substantially higher values of the Dzyaloshinskii-Moriya coupling.
  
    The article has the following structure. Section \ref{sec: Hamiltonian construction} is devoted to generalities: derivation of an electron Hamiltonian in the presence of a texture, and of its effective-mass (low-energy) limit, with the focus on a specific location of the band extrema. This section expands on the results of Ref. \cite{Davier_2023} -- both to make the presentation self-contained and to provide some additional details. 
  In Sec. \ref{sec: BP Profile, small energies} we use the low-energy Hamiltonian to treat shallow bound states for the Belavin-Polyakov skyrmion. The deep bound-state limit is next studied in Sec. \ref{sec: BP Deep bound states}, allowing us to complete the picture of the bound state evolution as a function of the BP skyrmion radius. In Sec. \ref{sec: Domain wall profile}, we address the bound-state problem for a DW skyrmion, exploring both shallow and deep states. In Sec. \ref{sec: Half filling energetics}, we estimate the contribution of bound states to the skyrmion energy as a function of the skyrmion shape. We also study electron states, bound to a straight domain wall, and evaluate their contribution to the domain-wall energy. 
  Which allows us to demonstrate how texture-bound electron  states substantially shift the transition line between the uniform and the modulated phases \cite{Bogdanov_1989}.
 The article concludes with a discussion in Sec. \ref{sec: Discussion}, while some of the technical details are given in the Appendices.

		\section{The Hamiltonian}
		\label{sec: Hamiltonian construction}
		
		\subsection{General arguments and the uniform state}
		\label{subsec:general}
  
		We start by writing down the effective Hamiltonian of a band electron in a collinear Néel antiferromagnet in the second-quantization formalism: 
		\begin{equation}
			\mathcal{H} = \int d^2r \; \Psi^\dagger (\mathbf{r}) \left[\varepsilon \left(\hat{\mathbf{p}}\right) + J_e\mathbf{M}(\mathbf{r})\cdot {\bm\sigma}\right]\Psi(\mathbf{r})
		\end{equation} 
	    with $\Psi^\dagger(\mathbf{r})$ and $\Psi (\mathbf{r})$ the electron creation/annihilation operators at point $\mathbf{r}$, and  $\varepsilon\left(\hat{\mathbf{p}}\right)$ the electron dispersion in the absence of Néel order. The second term $J_e\mathbf{M}(\mathbf{r})\cdot {\bm\sigma}$ accounts for the exchange interaction between the spontaneous local magnetization $\mathbf{M}(\mathbf{r})$ and the electron spin, represented by the triad of Pauli matrices ${\bm\sigma}$. Without loss of generality, hereafter we consider a square lattice with spacing $a_0$. N\'eel order can be defined via nearly opposite sublattice magnetizations $\mathbf{M}_1$ and $\mathbf{M}_2$, from which one can build the local 
 Néel order parameter $\mathbf{n}(\mathbf{r}) = \frac{\mathbf{M}_1 - \mathbf{M}_2}{2M_s}$ and the effective local magnetization $\mathbf{m}(\mathbf{r}) = \frac{ \mathbf{M}_1 + \mathbf{M}_2 }{2M_s}$, 
     where $M_s$ is the saturation magnetization. In the N\'eel state, the order parameter $\mathbf{n}(\mathbf{r})$ is large compared with the effective local magnetization: $\lVert\mathbf{n}\lVert \simeq 1 \gg \lVert\mathbf{m}\lVert$; moreover, $\mathbf{n}(\mathbf{r})$ varies smoothly in space and slowly in time, while $\mathbf{m}(\mathbf{r})$ fluctuates wildly. Therefore, hereafter we ignore the effective local magnetization $\mathbf{m}(\mathbf{r})$ altogether and approximate the magnetization by its dominant Fourier harmonic as per $\mathbf{M}(\mathbf{r}) \simeq M_s \mathbf{n}(\mathbf{r})  
     e^{i\mathbf{Q}\cdot\mathbf{r}}$ with $\mathbf{Q} = \left(\pm\frac{\pi}{a_0}, \pm\frac{\pi}{a_0}\right)$. 
 
\begin{figure}[ht]
	\centering \includegraphics[width=0.8\linewidth]{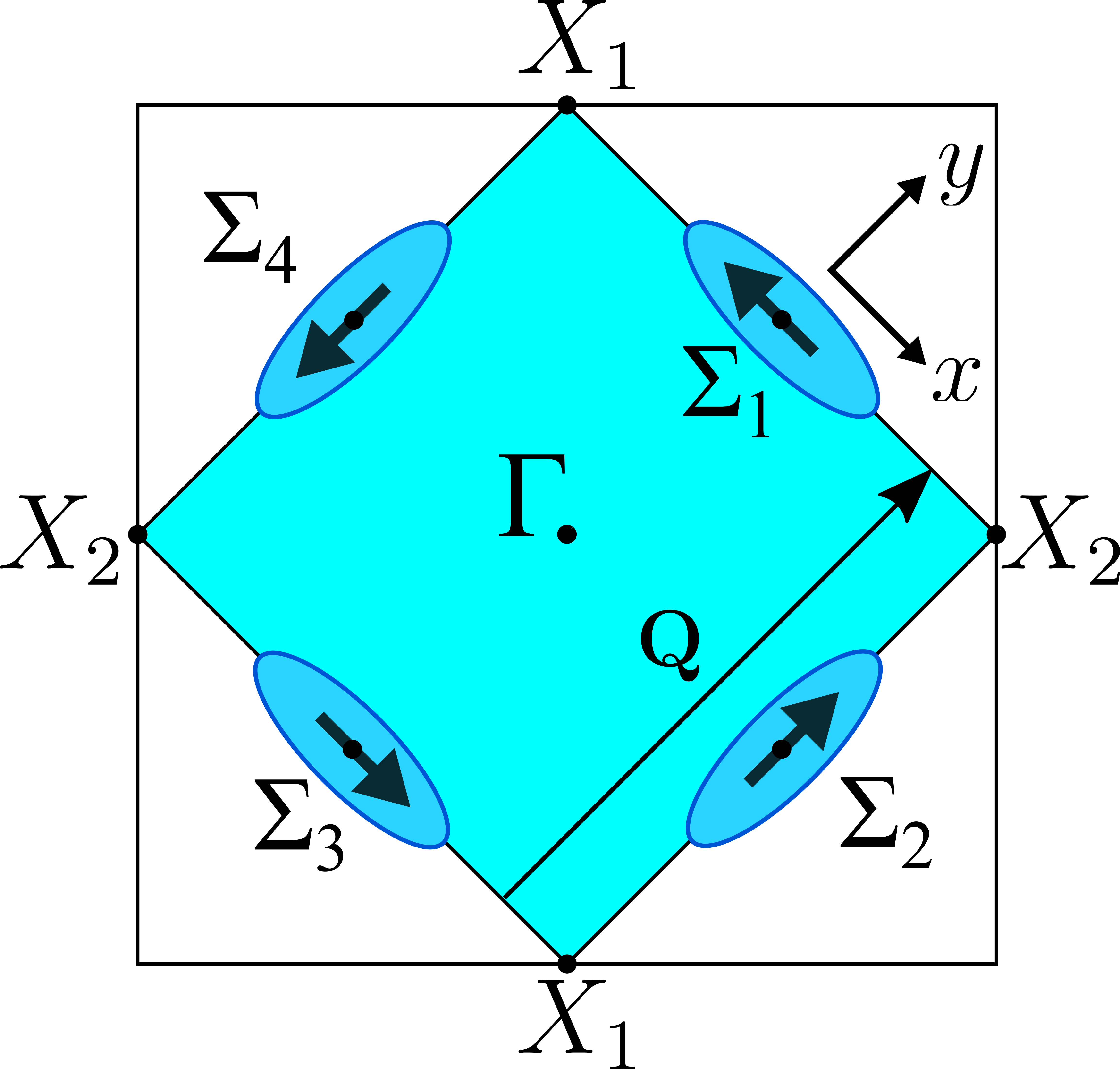}
	\caption{Magnetic Brillouin zone of the N\'eel state (cyan square), and  the paramagnetic Brillouin zone (white square). Points $\Sigma_i$ with small valleys, described by the expansion (\ref{Developpement Ep}), are shown symbolically in light blue. The bound sates at these valleys are spin-polarized, with polarization depending on the skyrmion profile. The polarization above corresponds to a N\'eel BP skyrmion, see main text for details. }
	\label{fig: MBZ}
\end{figure}

     It is convenient to begin with the uniform N\'eel state, and choose the magnetization to point along the $z$ axis in spin space: 
     $\mathbf{M}(\mathbf{r}) \simeq M_s \mathbf{e}_z 
     e^{i\mathbf{Q}\cdot\mathbf{r}}$. 
     Upon introducing the notation $\Delta = J_e M_s$, the Hamiltonian reads
	    \begin{equation}
	    	\mathcal{H} = \int d^2r \; \Psi^\dagger (\mathbf{r}) \left[\varepsilon \left(\hat{\mathbf{p}}\right) + \Delta \sigma_z 
      \, e^{i\mathbf{Q}\cdot\mathbf{r}}\right]\Psi(\mathbf{r}). \label{H_r_AF}
	    \end{equation} 
       Following transition to reciprocal space, the Hamiltonian takes the form
	    \begin{equation}
	    	\mathcal{H} = \sum_{\mathbf{p}\in \text{MBZ}} \begin{pmatrix}
	    		\Psi^\dagger_\mathbf{p} & \Psi^\dagger_\mathbf{p+Q}
	    	\end{pmatrix} 
	    	\begin{pmatrix}
	    		\varepsilon_\mathbf{p} & \Delta \sigma_z\\ 
	    		\Delta \sigma_z  & \varepsilon_\mathbf{p+Q}
	    	\end{pmatrix} 
	    	\begin{pmatrix}
	    		\Psi_\mathbf{p} \\ \Psi_\mathbf{p+Q} 
	    	\end{pmatrix}. \label{H_uniforme}
	    \end{equation} 
	  	That is, N\'eel order couples electron states at any two momenta $\mathbf{p}$ and $\mathbf{p+Q}$ \cite{Kulikov1984REVIEWSOT}, 
	  	reducing the Brillouin zone in the paramagnetic state to its half, as shown in Fig. \ref{fig: MBZ}.
    
  	The Hamiltonian is block-diagonal, and can be easily diagonalized to produce two doubly-degenerate electron bands
  		\begin{equation}
  			E_{\pm} = \varepsilon_+ \pm \sqrt{\Delta^2+\varepsilon_-^2}, \hspace{0.7cm} \varepsilon_\pm = \frac{\varepsilon_\mathbf{p} \pm \varepsilon_\mathbf{p+Q}}{2}. \label{Energies_bandes}
  		\end{equation} 
  	These bands are separated by a gap of width $2\Delta$, opening at crossing points $\varepsilon_\mathbf{p} = \varepsilon_\mathbf{p+Q}$. 
        The 
        N\'eel order thus turns a half-filled metal into an insulator. This elementary treatment of the uniform N\'eel state allows us to understand the overall electron band structure, and puts us in a position to study a non-uniform texture $\mathbf{n}(\mathbf{r})$. 
  		
		\subsection{A non-uniform texture}
            \label{subsec:non-uniform}
            
		Consider the Néel order parameter $\mathbf{n}(\mathbf{r})$, varying smoothly in space -- that is, on a scale large compared with $a_0$. Because of this nonuniformity, Hamiltonian (\ref{H_r_AF}) can no longer be easily diagonalized. However, we can gain insight by rendering $\mathbf{n}(\mathbf{r})$ uniform via a spin rotation $U(\mathbf{r})$ acting as per
		\begin{equation}
			U^\dagger(\mathbf{r}) \Big[ \mathbf{n}(\mathbf{r}) \cdot \bm{\sigma} \Big] U(\mathbf{r}) = \sigma^z \label{def U}.
		\end{equation} 
		Upon transformation $\Psi \rightarrow U^\dagger \Psi$, the Hamiltonian (\ref{H_r_AF}) reads
		\begin{equation}
			\mathcal{H} = \int d^2r \; \Psi^\dagger (\mathbf{r})  \left[U^\dagger(\mathbf{r})\varepsilon \left(\hat{\mathbf{p}}\right)U(\mathbf{r}) + \Delta \sigma^z e^{i\mathbf{Q}\cdot\mathbf{r}}\right] \Psi(\mathbf{r}). \nonumber
		\end{equation}
            The price for reducing the second term in the square brackets to that of Eq. (\ref{H_r_AF}) amounts to the appearance of a Peierls substitution \cite{Volovik1987} in the kinetic term, since  
		\begin{equation}
			 U^\dagger \hat{p}_i U = \hat{p}_i + \mathbf{A}_i\cdot \bm{\sigma} \label{Peierls}
		\end{equation}
		where $\mathbf{A}_i\cdot \bm{\sigma} = A_i^\alpha \sigma^\alpha = -i\hbar U^\dagger ( \partial_i U )$, with $i$ the spatial indices and $\alpha$ the spin indices (summation over repeated indices is implied). Various properties of this gauge field will be discussed later. Switching back to real space, the single-particle Hamiltonian, acting in the basis $\left(\Psi_\mathbf{p}, \Psi_{\mathbf{p+Q}}\right)$, reads
		\begin{equation}
			H = 
			\begin{pmatrix}
				\varepsilon_{\mathbf{p}}\left(\hat{\mathbf{p}} + \mathbf{A}\cdot\bm{\sigma}\right) & \Delta \sigma^z\\ 
				\Delta \sigma^z  & \varepsilon_{\mathbf{p+Q}}\left(\hat{\mathbf{p}} + \mathbf{A}\cdot\bm{\sigma}\right)
			\end{pmatrix} 
			\label{H_non_uniforme}
		\end{equation} 
  with $\varepsilon_{\mathbf{p}}\left(\hat{\mathbf{p}} + \mathbf{A}\cdot\bm{\sigma}\right) = U^\dagger \varepsilon\left(\hat{\mathbf{p}}\right)U$. 
  Because of the Peierls substitution, the terms on the diagonal in Eq. (\ref{H_non_uniforme}) are no longer proportional to the unit matrix in spin space. 

            With the band structure of the previous section in mind, and with Hamiltonian (\ref{H_non_uniforme}) at hand, we can now focus on the electron states near the band extrema.      
		
	\subsection{Low-energy effective Hamiltonian}
		
		We wish to derive an effective-mass Hamiltonian \cite{Kittel1963} describing an electron near the conduction band minimum at point $\mathbf{p}_0$. To this end, we replace the electron momentum $\mathbf{p}$ by $\hat{\mathbf{p}} = -i\hbar\bm{\nabla}$, and expand the dispersion  relations $\varepsilon(\mathbf{p})$ and $\varepsilon(\mathbf{p+Q})$, denoted as $\varepsilon_{\mathbf{p}_0}(\mathbf{p})$  and $\varepsilon_{\mathbf{p}_0 + \mathbf{Q}}(\mathbf{p})$, around the given conduction band minimum $\mathbf{p}_0$. Note that $\mathbf{p}_0$ must be a crossing point $\varepsilon(\mathbf{p}_0) = \varepsilon(\mathbf{p}_0+\mathbf{Q})$ of the uniform state. With the help of a unitary operator 
  \footnote{ Note that this operator is not unique since, at point $\mathbf{p}_0$, Hamiltonian (\ref{H_uniforme}) commutes with $\sigma^z$. This implies that $P$ is defined up to any two independent spin rotations around $\mathbf{e}_z$, for the conduction and the valence band, respectively. Eq. (\ref{matrice de passage}) corresponds to the simplest choice, with no spin rotation  around $\mathbf{e}_z$ in either of the two bands.} 
		\begin{equation}
    		P = \frac{1}{\sqrt{2}}\begin{pmatrix}
				1\!\!1 & 1\!\!1 \\
				\sigma^z & -\sigma^z 
			\end{pmatrix}
                \label{matrice de passage}
		\end{equation}
   we now switch to the basis where Hamiltonian (\ref{H_non_uniforme}) becomes diagonal at $\mathbf{p}_0$: 
		\begin{equation}
			H \to P^\dagger H P =  \begin{pmatrix}
				\hat{\bm{\gamma}}_+ + \Delta 1\!\!1  & \hat{\bm{\gamma}}_- \\ 
				\hat{\bm{\gamma}}_- & \hat{\bm{\gamma}}_+ - \Delta 1\!\!1 
			\end{pmatrix},  
			  \label{H complete 4x4}
		\end{equation}
		with 
		\begin{equation}
			\hat{\bm{\gamma}}_\pm (\hat{\mathbf{p}}) = \frac{1}{2} \left[\varepsilon_{\mathbf{p}_0}\left(\hat{\mathbf{p}} + \mathbf{A}\cdot\bm{\sigma}\right) \pm \sigma^z \varepsilon_\mathbf{p_0+Q}\left(\hat{\mathbf{p}} + \mathbf{A}\cdot\bm{\sigma}\right) \sigma^z\right].
            \label{eq:gamma_pm}
		\end{equation} 
  
 \begin{figure}
     \centering
     \includegraphics[width = \linewidth]{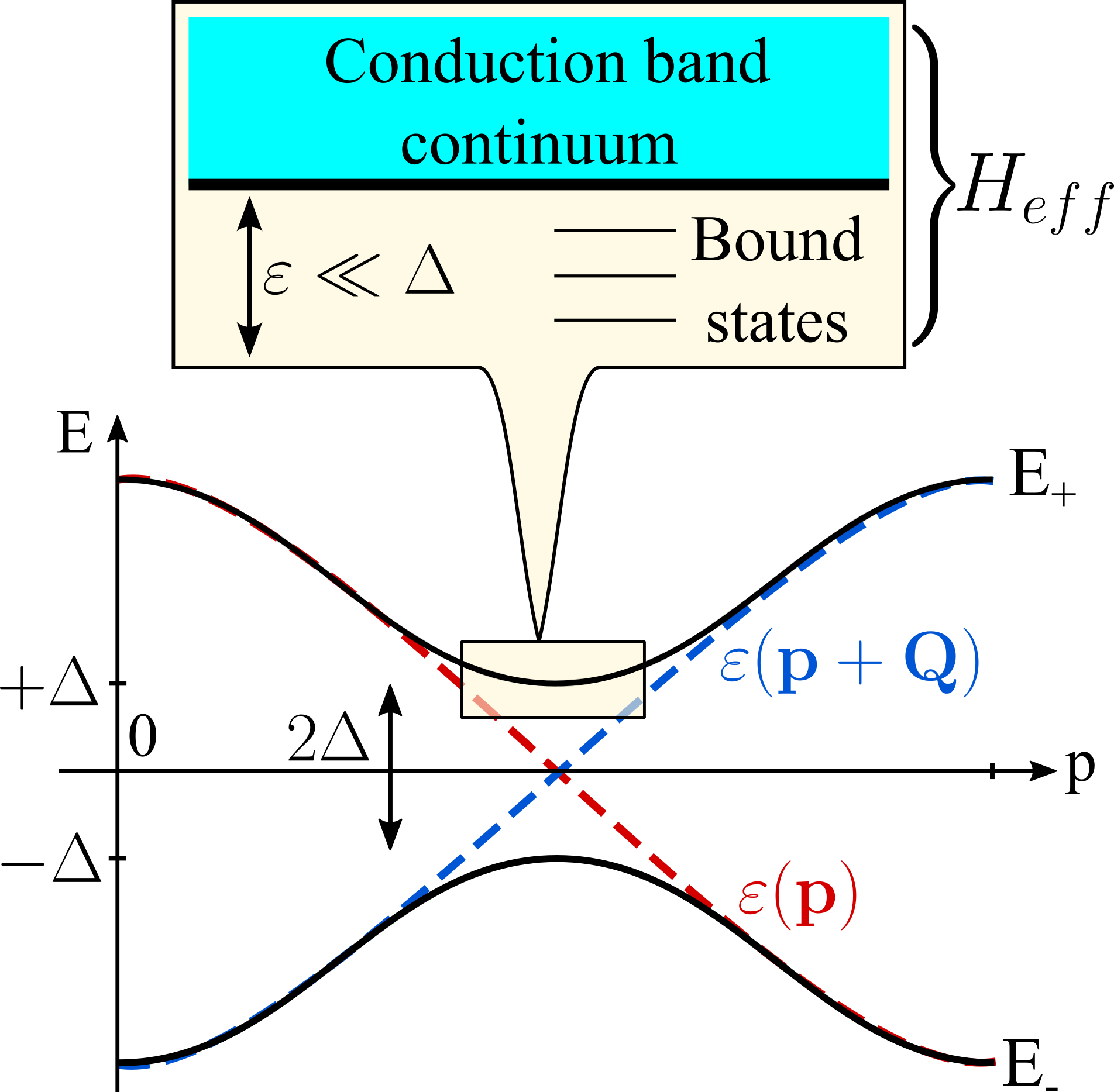}
     \caption{Deriving Hamiltonian $H_{eff}$ in Eq. (\ref{Hamiltonien_effectif}) from the full $4 \times 4$ 
     Hamiltonian of Eq. (\ref{H complete 4x4}). Hamiltonian (\ref{Hamiltonien_effectif}) provides an effective description near the band extremum.
     }
     \label{fig: Effective Hamiltonian development}
 \end{figure}
          
          From Hamiltonian (\ref{H complete 4x4}) acting on a bispinor, we now wish to construct an effective Hamiltonian, 
          that would describe only states near the bottom of conduction band (see Fig.~\ref{fig: Effective Hamiltonian development}), and thus would act only on a single spinor. This is akin to taking the Dirac Hamiltonian to its non-relativistic Pauli-Schr\"odinger limit \cite{Landau_vol4, Ryder1996Quantum}.
   To this end, we apply Hamiltonian (\ref{H complete 4x4}) to the bispinor $\left(\Phi, \chi\right)$. We then consider an electron at energy $E = \Delta + \varepsilon$ near the conduction band minimum $+\Delta$,  with $| \varepsilon | \ll \Delta$. This yields two equations 
		\begin{equation}
			\begin{split}
				\hat{\bm{\gamma}}_-\chi &= \left( \varepsilon - \hat{\bm{\gamma}}_+\right)\Phi,  \\
				\hat{\bm{\gamma}}_-\Phi &= \left(2\Delta + \varepsilon - \hat{\bm{\gamma}}_+ \right)\chi. \label{sous_eq_Schro}
			\end{split}
		\end{equation}
		Treating $\hat{\bm{\gamma}}_\pm (\hat{\mathbf{p}})$ as small against $\Delta$, we limit ourselves to leading terms of their momentum expansion near $\mathbf{p}_0$. To first order in $\frac{\hat{\bm{\gamma}}_-}{\Delta} \ll 1$, one finds $\chi \simeq \frac{\hat{\bm{\gamma}}_-}{2\Delta} \Phi$, which means $\Phi \gg \chi$. This allows us to write down the effective Schr\"odinger equation acting only on the spinor $\Phi$:  
		\begin{equation}
			\left(\hat{\bm{\gamma}}_+ + \frac{\hat{\bm{\gamma}}_-^2}{2\Delta} \right) \Phi = \varepsilon \Phi,
		\end{equation}
		which yields the sought effective Hamiltonian to first order in $\frac{\varepsilon}{\Delta} \ll 1$:
		\begin{equation}
			H_{eff} = \hat{\bm{\gamma}}_+ + \frac{\hat{\bm{\gamma}}_-^2}{2\Delta} .
                \label{Hamiltonien_effectif}
		\end{equation} 	
The leading terms in the momentum expansion of $\hat{\bm{\gamma}}_\pm$ depend on the location of the minimum $\mathbf{p}_0$ in the Brillouin zone. Here we focus on points $\Sigma_i$: the face centers of the magnetic Brillouin zone in Fig. \ref{fig: MBZ}. The peculiarity of these points is the presence of a linear term in the momentum expansion of $\varepsilon_\mathbf{p_0} \left(\mathbf{p}\right)$ 
and $\varepsilon_\mathbf{p_0+Q}\left(\mathbf{p}\right)$: 
\begin{equation}
		\begin{split}
			&\varepsilon_\mathbf{p_0}\left(\mathbf{p}\right) =  v p_y + \frac{p_i^2}{2m_i} + \mathcal{O}\left(\left\|\mathbf{p}^3\right\|\right),\\ \label{Developpement Ep}
			&\varepsilon_\mathbf{p_0+Q}\left(\mathbf{p}\right) =  - v p_y + \frac{p_i^2}{2m_i} + \mathcal{O}\left(\left\|\mathbf{p}^3\right\|\right).
		\end{split}
\end{equation} 
Here $v$ is the Fermi velocity of the parent paramagnetic state along the $\hat{y}$ direction, and $m_i$ are the effective masses along $\hat{x}$ and $\hat{y}$,  as shown in Fig. \ref{fig: MBZ}. 

Equations (\ref{Developpement Ep}) yield the expansion of $\hat{\bm{\gamma}}_\pm (\hat{\mathbf{p}})$ in Eq. (\ref{eq:gamma_pm}) to second order in momentum:
\begin{equation}
	\begin{split} 
		&\hat{\bm{\gamma}}_+ (\hat{\mathbf{p}}) = v  \mathbf{A}_{y}^\parallel\cdot\bm{\sigma} + \frac{1}{2m_i}\left[\left(\hat{p}_i + A_i^z\sigma^z\right)^2 + \left( A_i^\parallel\right)^2\right], \\
		&\hat{\bm{\gamma}}_- (\hat{\mathbf{p}}) = v (\hat{p}_y+A_{y}^z\sigma^z) + \frac{1}{2m_i}\{\hat{p}_i,\mathbf{A}_i^\parallel \cdot\bm{\sigma} \},
	\end{split} \label{gammas}
\end{equation}
 where $\mathbf{A}_i^\parallel= (A_i^x, A_i^y,0)^t$ are the gauge-field components in the $xy$-plane in spin space. 
 Which, in turn, allows us to rewrite Hamiltonian (\ref{Hamiltonien_effectif}) to second order in momentum:
\begin{equation}
H = \frac{1}{2m_i^*}\left(\hat{p}_i + A_i^z\sigma^z\right)^2 
 + v \mathbf{A}_{y}^\parallel\cdot\bm{\sigma}
  + \frac{\left( A_i^\parallel\right)^2}{2 m_i} , 
  \label{H_01}
\end{equation}
where the second term in Eq. (\ref{Hamiltonien_effectif}) renormalizes the effective mass 
$m_y^*$ relative to its band value in the expansion (\ref{Developpement Ep}) as per 
\begin{equation}
	m_y^{-1} \;\longrightarrow \; (m_y^*)^{-1} = m_y^{-1} + \frac{v^2}{\Delta} ,
 \label{eq:anisotropy}
\end{equation}
with $\frac{m_y^*}{m_y} \sim \frac{\Delta}{\varepsilon_F} \ll 1$. As we will see below, it is thanks to 
this renormalization that the kinetic term $\frac{1}{2m_y^*}\left(\hat{p}_y + A_y^z\sigma^z\right)^2$ 
becomes comparable with the dominant potential term $v \mathbf{A}_{y}^\parallel\cdot\bm{\sigma}$ 
at smaller texture length scales $a \ll L \ll \xi$. 

 Hamiltonian (\ref{H_01}) admits further simplification: within continuum description, 
 the N\'eel order parameter $\mathbf{n}$ of an antiferromagnetic texture varies on a 
 length scale $L$, that is large compared with the lattice spacing $a_0$. The latter 
 is of the order of the `effective lattice spacing' $a \equiv \frac{\hbar}{m_y v}$. 
 As the reader will see below, the two key length scales in the problem are $a$ and the `coherence length' $\xi \equiv \frac{\hbar v}{\Delta}$. Their ratio is $\frac{a}{\xi} = \frac{\Delta}{\varepsilon_F} \ll 1$, with $\varepsilon_F \equiv m_yv^2$ the characteristic electron bandwidth. 
 For an electron wave function varying on the characteristic length scale $L_\psi \gg a$, we find 
\begin{equation}
 v A_i^\alpha 
 \sim \Delta \frac{\xi}{L}
 \gg 
 \frac{\left(A_i^\alpha\right)^2}{2m_i} , 
 \frac{\{\hat{p}_i,A_i^\alpha \} }{2m_i}  
 \sim
 \Delta \left\{ \frac{\xi a}{L^2} , \frac{\xi a}{L L_\psi} \right\}  
 \label{orders of magnitude} .
\end{equation} 
As a result, the last term in Eq. (\ref{H_01}) can be safely omitted at all relevant texture length scales below, 
and the leading terms of Hamiltonian (\ref{Hamiltonien_effectif}) to second order in momentum thus read 
\begin{equation}
	H = \frac{1}{2m_i^*}\left(\hat{p}_i + A_i^z\sigma^z\right)^2 
 + v \mathbf{A}_{y}^\parallel\cdot\bm{\sigma} .
 \label{H_0} 
\end{equation}
Notice that the effective mass $m_i^*$ above is substantially anisotropic: $m_x^* = m_x$ is of the order of 
the electron band mass or greater (on a square lattice with only the nearest-neighbour hopping, $m_x$ is infinite), 
and thus $m_x^* \gg m_y^*$. While mass anisotropy at points $\Sigma$ is generic as dictated by their position at the 
magnetic Brillouin zone boundary, strong mass anisotropy in Hamiltonian (\ref{H_0}) can be traced back to separation 
of scales: antiferromagnetic gap $\Delta$ in the electron spectrum being small relative to the band width $\varepsilon_F$.

As we show below, the second term $v \mathbf{A}_{y}^\parallel\cdot\bm{\sigma}$ of Hamiltonian (\ref{H_0}) tends to produce electron states, bound to topological textures such as skyrmions and domain walls. On the one hand, in a N\'eel antiferromagnet, such a term owes its existence to lower symmetry of band extrema $\Sigma_i$ in the Brillouin zone: at higher-symmetry extrema (e.g. points $\Gamma$ and $X$ in Fig.~\ref{fig: MBZ}), such a term is not allowed. 
On the other hand, in a ferromagnet, such a term is not allowed at all, which underlines a key difference between 
ferro- and antiferromagnets. We briefly compare the two problems in Appendix \ref{Appendix : comparison with FM}.

Hamiltonian (\ref{H_0}) may appear odd 
to a reader used to working with electromagnetic field. 
Thus we now take a closer look at transformation properties of the gauge field $\mathbf{A}$.

\subsection{Gauge transformations}
\label{subsec:transformations}
		
Unitary operator $U(\mathbf{r})$ of Eq. (\ref{def U}) is defined up to a spin rotation $V_z(\mathbf{r})=e^{i\chi(\mathbf{r})\sigma^z}$ around the $z$ axis, with a single-valued field $\chi(\mathbf{r})$. Indeed, such a rotation leaves Eq. (\ref{def U}) intact: 
\begin{equation}
	V_z^\dagger U^\dagger \Big[ \mathbf{n} \cdot \bm{\sigma} \Big] UV_z =  V_z^\dagger \sigma^z V_z = \sigma^z ,
\end{equation} 
which means that $V_z(\mathbf{r})$ amounts to a gauge transformation. Under such a transformation, $\mathbf{A}_i\cdot \bm{\sigma}$ of Eq. (\ref{Peierls}) varies  as per 
\begin{equation}
	\begin{split}
		\tilde{\mathbf{A}_i} \cdot \bm{\sigma}  &= -i\hbar V_z^\dagger U^\dagger \partial_i \left(U V_z\right)\\ 
            &= V_z^\dagger \mathbf{A}_i\cdot \bm{\sigma} V_z -i\hbar V_z^\dagger \partial_i V_z \\ \label{transformation de jauge}
		&= V_z^\dagger \mathbf{A}_i^\parallel \cdot \bm{\sigma} V_z + A_i^z\sigma^z + \hbar \partial_i\chi \sigma^z.
	\end{split}
\end{equation}	
That is, the $A_i^z$ component transforms as electromagnetic vector potential, while the transverse component $\mathbf{A}_i^\parallel = ( A_i^x , A_i^y,0)^t$ 
simply rotates by $2\chi (\mathbf{r})$ around the $z$ axis in spin space. While the different spin components $\mathbf{A}_i^\alpha \sigma^\alpha$ do not commute, they split into $A_i^z \sigma^z$ and $\mathbf{A}_i^\parallel \cdot \bm{\sigma}$, each obeying its distinct abelian transformation law. 

Another transformation that leaves Eq. (\ref{def U}) invariant is a spin rotation $W(\mathbf{r})= e^{i\phi(\mathbf{r})\mathbf{n}\cdot \bm{\sigma}}$ around the local $\mathbf{n}(\mathbf{r})$: 
\begin{equation} 
\label{transformation_W}
	 U^\dagger W^\dagger \Big[ \mathbf{n} \cdot \bm{\sigma} \Big] WU =  U^\dagger \mathbf{n} \cdot \bm{\sigma} U = \sigma^z ,
\end{equation} 
which implies gauge transformation 
\begin{equation}
	\tilde{\mathbf{A}_i} \cdot \bm{\sigma} = \mathbf{A}_i\cdot \bm{\sigma} - i\hbar U^\dagger \left[W^\dagger \partial_i W\right] U. \label{transformation de jauge 2}
\end{equation} 
Just as for transformation $V_z$ above, a simple calculations shows that, under $W$, $\mathbf{A}_i^{\|}$ undergoes an in-plane rotation by $2 \phi$, while $A_i^z$ transforms as electromagnetic vector potential (see Appendix \ref{Appendix : Gauge transformation}). 

Note that, for infinitesimal $\chi (\mathbf{r})$ and $\phi (\mathbf{r})$, the relations (\ref{transformation de jauge}) and (\ref{transformation de jauge 2}) read, respectively 
\begin{equation}
    \delta \mathbf{A}_i\cdot\bm{\sigma} = i\hbar \chi [\mathbf{A}_i\cdot\bm{\sigma}, \sigma^z] + \hbar \partial_i\chi \sigma^z
\end{equation}
and
\footnote{Here we use the identity 
\begin{equation*}
   - i\hbar \partial_i \left( U^\dagger \mathbf{n}\cdot \bm{\sigma} U\right) = [\sigma^z , \mathbf{A}_i \cdot \bm{\sigma}] - i\hbar U^\dagger \partial_i\mathbf{n}\cdot \bm{\sigma} U = 0.
\end{equation*}
}
\begin{equation}
    \begin{split}
        \delta \mathbf{A}_i\cdot\bm{\sigma} &= \hbar \phi \, U^\dagger [\partial_i\mathbf{n}\cdot \bm{\sigma}] U + \hbar \partial_i\phi \sigma^z \\
        &= i\hbar \phi \, [\mathbf{A}_i\cdot\bm{\sigma}, \sigma^z] + \hbar \partial_i\phi \sigma^z 
    \end{split}
\end{equation}
and thus become identical. Invariant under these two transformations, Hamiltonians (\ref{H_01}) and (\ref{H_0}) enjoy the peculiar gauge symmetry above. 

To perform calculations for a given texture, we will have to choose a concrete $U(\mathbf{r})$ -- that is, to fix a gauge. 
We will define $\mathbf{n}$ via its polar angle $\theta$ and azimuthal angle $\phi$ as per 
\begin{equation}
\mathbf{n} = \left(\sin\theta \cos\phi, \sin \theta \sin \phi , \cos\theta\right)^t ,
\label{eq:n}
\end{equation}
and choose \cite{Tatara_2008}
\begin{equation} 
U(\mathbf{r}) = \mathbf{m}\cdot \bm{\sigma} 
\label{eq:m}
\end{equation} 
with $\mathbf{m}$ being the bisector between $\mathbf{n}$ and $\mathbf{e}_z$. 
That is, $\mathbf{m}\cdot \bm{\sigma}$ amounts to a $\pi$-rotation around $\mathbf{m}$. Such a choice makes calculations straightforward with the help of the textbook Pauli-matrix property $(\mathbf{n}\cdot\bm{\sigma}) \,(\mathbf{m}\cdot\bm{\sigma}) = \mathbf{n}\cdot\mathbf{m} + i\, (\mathbf{n}\times \mathbf{m})\cdot\bm{\sigma}$, yielding
\begin{equation}
	\mathbf{A}_i\cdot\bm{\sigma} = -i \hbar\, (\mathbf{m}\cdot \bm{\sigma})\, (\partial_i \mathbf{m}\cdot \bm{\sigma}) = \hbar\, (\mathbf{m} \times \partial_i \mathbf{m}) \cdot \bm{\sigma}.
\end{equation}
A simple calculation leads to the following expression:  
\begin{equation}
	\hbar^{-1}\mathbf{A}_i =  \frac{\partial_i\theta}{2} \begin{pmatrix}
		-\sin\phi \\
		\cos \phi \\
		0
	\end{pmatrix}
	+  \frac{\partial_i \phi }{2} \begin{pmatrix}
		-\sin\theta \cos\phi \\
		-\sin\theta \sin\phi \\
		1-\cos\theta
	\end{pmatrix}. \label{definition A_i}
\end{equation} 

As with electromagnetic field, certain combinations of the $\mathbf{A}_i$ components and their gradients prove gauge-invariant. In appendix \ref{Appendix : Gauge invariants}, we show that these can be expressed via the gradient energy density 
$({\bm \nabla} {\bf n})^2$ and the the skyrmion number density $({\bf n} \cdot \left[ \partial_x {\bf n} \times \partial_y {\bf n} \right])$.

Now that we constructed an effective-mass Hamiltonian for a generic antiferromagnetic texture, and elucidated its gauge-transformation properties, we are in a position to study an electron in the presence of a skyrmion.

\section{Belavin-Polyakov profile, shallow bound states}
\label{sec: BP Profile, small energies}
\subsection{Profile properties}
\label{subsec: Profile properties}

Note that a skyrmion has its own localized eigenmodes that, generally, must be considered on an equal footing with the electron degrees of freedom. There is, however, a realistic limit, where proper frequencies of the skyrmion prove to be small relative to those of the electron motion. Then, to first approximation, the texture can be treated as static input to the electron problem. Once the electron eigenstates are found, a comparison of characteristic electron frequencies with those of the proper modes of the skyrmion allows one to define the validity range of such a calculation, as it was done in our prior work for a skyrmion close to the Belavin-Polyakov (BP) profile \cite{Davier_2023}.

The BP skyrmion  naturally emerges in a perfectly isotropic magnet, described by the continuum-limit energy density $J\left(\bm{\nabla}\cdot \mathbf{n}\right)^2$ with stiffness $J$. In a topological sector labelled by the winding number $Q = \pm 1, \pm 2, \pm 3, ...$, the lowest-energy solution is the BP skyrmion of radius $R$, with $R$-independent energy $4\pi J | Q |$ \cite{Belavin_Polyakov_1975,Rajaraman}. 

The profile of a skyrmion is specified by the dependence of polar and azimuthal angles $\theta$ and $\phi$ in Eq. (\ref{eq:n}) 
on the coordinates in the plane. For the latter, we use polar coordinates $r = \sqrt{x^2+y^2}$ and $\alpha = \arctan\frac{y}{x}$. 
For high-symmetry skyrmion textures such as the BP skyrmion, the polar angle $\theta$ depends only on the radial coordinate $r$ in the plane, while the azimuthal angle $\phi$ depends on angle $\alpha$ in the plane as per $\phi = Q \alpha + \gamma$. The offset $\gamma$ is called helicity, while the coefficient $Q$ is the skyrmion winding number \cite{Sky_Gobel_2021}. 
A skyrmion with  $\gamma = 0$ is often called \textit{N\'eel} skyrmion, while its 
counterpart with $\gamma = \frac{\pi}{2}$ is referred to as \textit{Bloch} skyrmion.

Here we limit ourselves to $Q=1$, and consider the BP profile defined as per \cite{Belavin_Polyakov_1975,Rajaraman}
\begin{equation}
	\sin \theta = \frac{2z}{1 + z^2}, \hspace{1cm} z = \frac{r}{R} \label{def theta BP},
\end{equation}
depending on a single length scale, the skyrmion radius $R$. At first sight, the definition (\ref{def theta BP}) does not imply a choice of boundary conditions for the orbital angle $\theta(r)$. Furthermore, for an antiferromagnet, the choice of boundary conditions should be  physically irrelevant since it relies on an arbitrary choice of definition for the Néel order parameter $\mathbf{n} = \pm \frac{\mathbf{M}_1-\mathbf{M}_2}{2M_s}$. However, when looking for example at the $z$ components of the gauge field 
\begin{equation}
	A_i^z = \hbar\frac{\partial_i\phi}{2}\left(1-\cos\theta\right) ,
\end{equation}
we see that the choice $\theta(0) = \pi$ makes the origin singular. The only possible choice of boundary conditions is thus $\theta(0) = 0$ and $\lim_{r\to\infty}\theta(r) = \pi$, which allows to obtain $\cos\theta = 2 \frac{1-z^2}{1+z^2}$, leading to 
\begin{equation}
	\label{eq:Az}
	A^z_x = \hbar\frac{- y}{R^2 + r^2} \,\, , \hspace{1cm} A^z_y = \hbar\frac{x}{R^2 + r^2} .
\end{equation}
Encircling the skyrmion along a large closed contour makes the electron accumulate a geometric flux of $2\pi$: 
$\sigma^z \oint A^z_i dl_i = 2\pi\hbar \sigma^z$, which gives rise to the skyrmion Hall Effect \cite{TSHE_Yin_2015}. 
This is a consequence of a more general relation (\ref{invariant de jauge}): an electron encircling a skyrmion of topological charge $Q$  accumulates a geometric flux $2\pi Q$, independent of the skyrmion profile.\\

The BP profile has a remarkable property that greatly simplifies the problem. 
To illustrate it, consider the expression for the $A_y^x$, produced by a 
high-symmetry $Q=1$ skyrmion with $\phi = \alpha + \gamma$:
\begin{equation}
	\begin{split}
		A_y^x &= -\frac{\hbar}{2}\cos\alpha \sin\alpha\sin\gamma \left(\frac{\sin\theta}{r}-\frac{d\theta}{dr}\right)\\ 
		&- \frac{\hbar}{2}\cos\gamma \left(\cos^2\alpha \frac{d\theta}{dr} + \sin^2\alpha \frac{\sin\theta}{r} \right) . \label{Expression_explicit_A_y^x}
	\end{split}
\end{equation}
According to Eq. (\ref{Expression_explicit_A_y^x}), variables $r$ and $\alpha$ separate for a profile, 
satisfying 
\begin{equation}
	\frac{d\theta}{dr} = \frac{\sin\theta}{r} .
\end{equation}
Remarkably, this very relation is obeyed by the $Q=1$ BP profile \cite{John_1995}, implying 
\begin{equation}
	\mathbf{A}_y^\parallel = - \hbar \frac{\sin\theta}{2r} \mathbf{e}_\sigma , 
\end{equation} 
where unit vector $\mathbf{e}_\sigma = (\cos\gamma, \sin \gamma, 0)^t$ defines the orientation of $\mathbf{A}_y^\parallel$ in spin space. 
Note that $\mathbf{e}_\sigma$ only depends on the helicity, and remains uniform in real space. Without loss of generality, this allows us to limit ourselves to a N\'eel skyrmion  
($\gamma = 0$), for which $\mathbf{e}_\sigma = \mathbf{e}_x$ 
\footnote{ For a $Q \neq 1$ skyrmion with $\phi = Q\alpha + \gamma$, the BP profile can be defined via 
\begin{equation*}
	 \frac{d\theta}{dr} = Q\frac{\sin\theta}{r} .
\end{equation*}
The dominant potential term in Eq.~\ref{H_0} then reads 
\begin{equation*}
	v\mathbf{A}_y^\|\cdot \bm{\sigma} =   \frac{\xi}{R}\frac{Q \Delta}{1+z^2} \, \mathbf{e}_\sigma^Q \cdot \bm{\sigma} ,
\end{equation*}
where $\mathbf{e}_\sigma^Q = (\cos \phi, \sin \phi,0)^t$, and $\phi \equiv \gamma + (Q-1)\alpha$. Thus, for $Q \neq 1$, 
orientation of $\mathbf{A}_y^\|$ in spin space becomes non-uniform. 
}.

We will now study shallow electron states, bound to the BP skyrmion as described by Hamiltonian (\ref{H_0}).
 
\subsection{Shallow bound states}
\label{subsec: Small energies}

The first thing to note is that strong anisotropy $m_x^* \gg m_y^*$ makes $y$ a `fast' variable relative to $x$, which calls for the Born-Oppenheimer approximation \cite{Tully2001}. Recalling that $m_y^* \sim \frac{\Delta}{v^2}$, Hamiltonian (\ref{H_0}) can be recast as
\begin{equation}
\begin{split}
		\frac{H}{\Delta} = \frac{\xi^2}{2R^2} \left(-\frac{\partial^2}{\partial \tilde{y}^2} 
		+ \frac{\tilde{x}^2}{(1+z^2)^2}\right)
		- \frac{\xi}{R} \frac{1}{1+z^2} \sigma^x ,
\end{split}
\end{equation}
with $\tilde{x}_i = \frac{x_i}{R}$. Equation above allows us to demonstrate the appearance of skyrmion-electron bound states, polarized along $\mathbf{e}_x$. 

Above, anticommutator $\left\{\hat{p}_y, A_y^z\sigma^z\right\}$ has been omitted as proportional to $\sigma^z$, and thus only acting in second order of perturbation theory, the dominant term being proportional to $\sigma^x$. This can be seen directly from the physics picture in Fig. \ref{fig: Structure interne bande}: the term $\left\{\hat{p}_y, A_y^z\sigma^z\right\}$ has matrix elements only across the gap, hence its contribution is suppressed by large perturbation-theory denominator $\Delta$. 

\begin{figure}
    \centering
    \includegraphics[width = \linewidth]{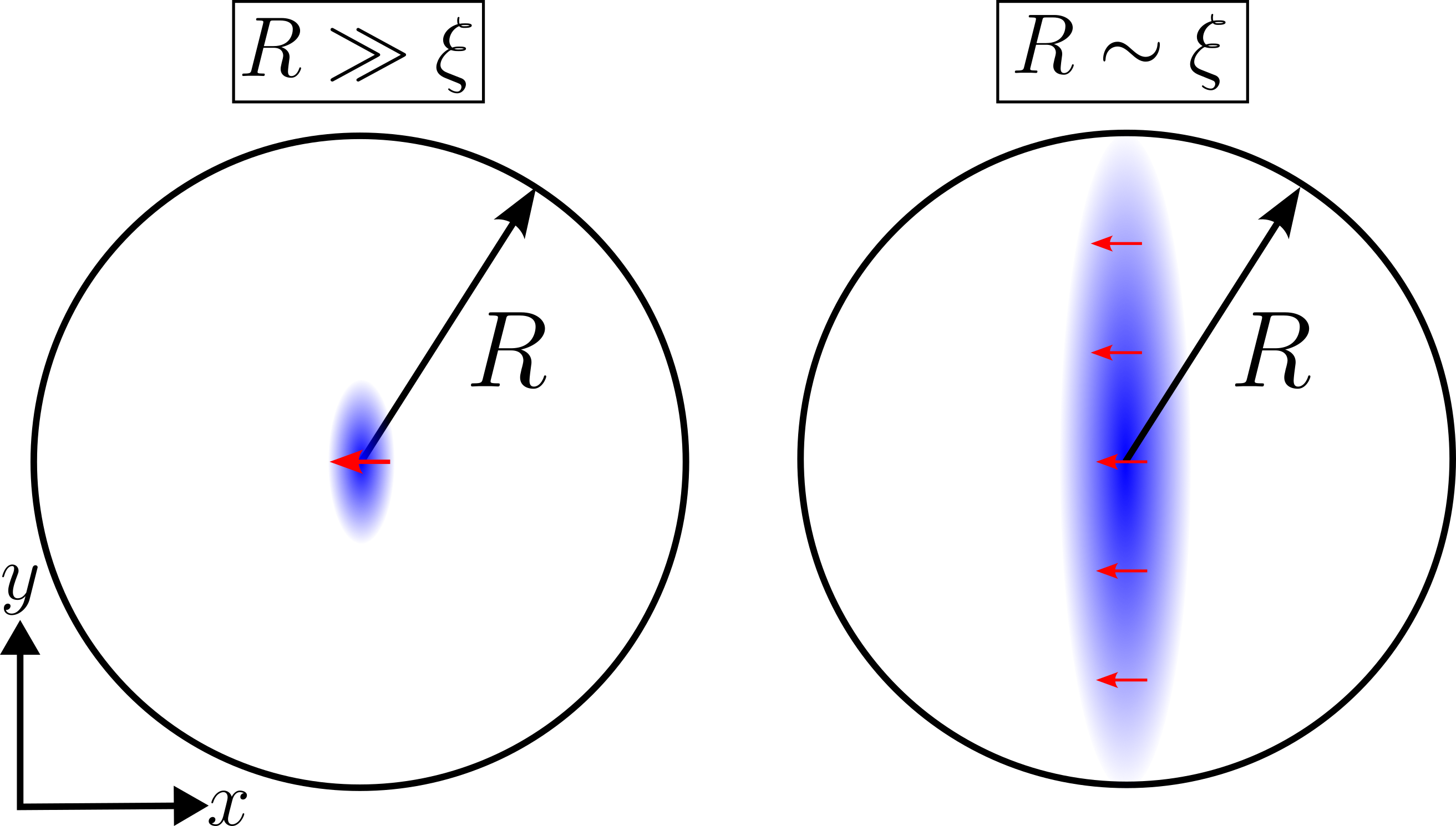}
    \caption{
    Density plot of the ground-state wave function (see Eq. (\ref{Energy_OH_BP})) of an electron, bound to a Belavin-Polyakov skyrmion of radius $R$, 
    for a large ($R\gg \xi$) and intermediate-size ($R \sim \xi$) skyrmion. For $R \gg \xi$, the wave function is localized well within the skyrmion core. It starts spreading beyond the skyrmion core as the skyrmion radius $R$ decreases to become of the order of the coherence length $\xi$, before spilling over $R$ for $R \lesssim \xi$ (not shown here). For a large ($R \gtrsim \xi$) N\'eel skyrmion, the low-lying bound states are those of an anisotropic harmonic oscillator. They are spin-polarized essentially uniformly along $-\mathbf{e}_x$, as shown in Fig. \ref{fig: MBZ}, as sketched by red arrows here. Along $y$, the characteristic length scale of the ground-state wave function is $\bar{y} \sim R \sqrt[4]{\frac{\xi}{R}}$, while the 
    scale along $x$ is $\bar{x} \sim R \sqrt[4]{\epsilon \frac{\xi}{R}} \ll \bar{y}$: 
    it remains small compared with $R$ even for $R \sim \xi$.}
    \label{fig: wave function BP}
\end{figure}

Considering the $|+\rangle_x$ component of the wave function, the corresponding low-lying states can be evaluated by expanding the potential around the bottom of the well. This yields the effective harmonic potential for the `slow' coordinate $\tilde{x} = \frac{x}{R}$:
\begin{equation}
	\frac{\varepsilon^n_x}{\Delta} = \frac{\xi}{R} \left(-1 + \left(n+\frac{1}{2}\right)\sqrt{\frac{2\xi}{R}}  + \left[1+\frac{\xi}{2R}\right]\tilde{x}^2\right).
\end{equation} 
Reintroducing the kinetic energy term $\frac{\hat{p}_x^2}{2m_x^*}$ then yields the low-lying states
\begin{equation}
	\begin{split}
		\frac{\varepsilon^n_m}{\Delta} &= \frac{\xi}{R} \left[-1 + \left(n+\frac{1}{2}\right)\sqrt{\frac{2\xi}{R}} \right] \\
		 &\hspace{1cm}+ \left(m+\frac{1}{2}\right) \sqrt{2\epsilon} \left(\frac{\xi}{R}\right)^{3/2} \sqrt{1+\frac{\xi}{2R}}
		 \label{Energy_OH_BP}
	\end{split}
\end{equation} 
with $\epsilon = \frac{m_y}{m_x}\frac{a}{\xi} \lll 1$. Parameter $\epsilon$ governs the ratio $\frac{\omega_x}{\omega_y}$ of the characteristic frequencies of the `slow' and `fast' degrees of freedom. The wave functions of these states are those of a two dimensional anisotropic harmonic oscillator with different frequencies along directions $x$ and $y$. These wave functions are thus localized at the skyrmion core, as shown in Fig.~\ref{fig: wave function BP} for the ground state (energy $\varepsilon_0^0$ in Eq. (\ref{Energy_OH_BP})).

The bound states remain shallow, and the harmonic expansion continues to hold for radia $R \gg \xi$. 
With $R$ approaching $\xi$, the frequency $\hbar\omega_y$ becomes comparable to $\Delta$, and the 
low-energy Hamiltonian (\ref{H_0}) breaks down. That is, for radia $R \lesssim \xi$, 
the complete Hamiltonian (\ref{H complete 4x4}) must be considered. \\

At the same time, with $R$ decreasing further below $\xi$, the repulsive term 
$\frac{\left(A_y^z\right)^2}{2 m_y^*} \sim \Delta\frac{\xi^2}{R^2}$ in the low-energy 
Hamiltonian (\ref{H_0}) grows relative to the attraction $v A_y \sim \Delta \frac{\xi}{R}$. 
Using approximation $A_y^z \sim \hbar \frac{x}{R^2}$ and estimating $\langle \hat{x} ^2 \rangle$ 
within harmonic approximation shows that the repulsion compensates the attraction $vA_y^x$ 
for $R \sim \sqrt{\epsilon} \xi$. (Notice that $\sqrt{\epsilon} \xi \gg a$, 
hence at this scale the continuum description of the texture still applies.) 
Therefore, bound states become shallow again and disappear 
at an $R = \bar{R} \sim \sqrt{\epsilon} \xi \ll \xi$. 
This estimate can be obtained more rigorously \cite{Davier_2023}, and does not rely 
on strong mass anisotropy. For a perfectly isotropic toy model, our calculation of 
$\bar{R}$ in Appendix \ref{Appendix: Isotropic toy model} yields the same result. 

We now turn to a study of deep bound states with the help of the full Hamiltonian (\ref{H complete 4x4}).

\section{Belavin-Polyakov profile: Deep bound states}
\label{sec: BP Deep bound states}

Above and in Ref. \cite{Davier_2023}, we have analysed shallow bound states at both small $R \gtrsim \bar{R}$ and large BP skyrmion radia $R \gg \xi$. Now we turn to  intermediate radia $\bar{R} \lesssim R \lesssim \xi$, where the bound-state energy becomes comparable with the gap $\Delta$.

\subsection{Particle-hole symmetry}
\label{subsec: Electron-hole symmetry}
Before turning to deep bound states, we would like to highlight the electron-hole symmetry of Hamiltonian (\ref{H complete 4x4}). If we go through steps (\ref{sous_eq_Schro}) to (\ref{Hamiltonien_effectif}), now considering energies near the valence band maximum and setting $E = -\Delta + \varepsilon$ with $| \varepsilon | \ll \Delta$, this time we obtain
\begin{equation}
	\Phi \simeq -\frac{\hat{\bm{\gamma}}_-}{2\Delta} \chi .
\end{equation}
The full 4-spinor wave function $(\phi , \chi)$ is now dominated by $\chi$. To first order in $\frac{\hat{\bm{\gamma}}_-}{\Delta} \ll 1$, the effective Hamiltonian for $\chi$ reads 
\begin{equation}
	H_- = \hat{\bm{\gamma}}_+ - \frac{\hat{\bm{\gamma}}_-^2}{2\Delta} \label{H basse énergie}.
\end{equation}  
Comparing Eqs. (\ref{Hamiltonien_effectif}) and (\ref{H basse énergie}), we see that shallow bound states near the top of the valence band have polarization $|-\rangle_{\mathbf{e}_\sigma}$ and energy $\varepsilon = E + \Delta > 0$, while their conduction-band counterparts have polarization $|+\rangle_{\mathbf{e}_\sigma}$ and energy $\varepsilon = E - \Delta < 0$. Both for large radia $R \gg \xi$ and for $R \gtrsim \bar{R}$, shallow bound states near the bottom of the conduction band and and their counterparts near the top of the valence bands obey the same Schr\"odinger equation, and their energies are perfectly symmetric relative to the gap center $E=0$. 
In other words, shallow bound states near the edge of the conduction band are related to their valence-band partners by particle-hole symmetry 
\begin{equation}
	|+\rangle_{\mathbf{e}_\sigma} \to |-\rangle_{\mathbf{e}_\sigma}, \hspace{1cm} \varepsilon \to - \varepsilon.
 \label{eq: particle-hole symmetry}
\end{equation} 
  
In the intermediate range $R \sim \xi$, the full $4 \times 4$ Hamiltonian (\ref{H complete 4x4}) has to be considered. 
Particle-hole symmetry (\ref{eq: particle-hole symmetry}) corresponds to wave function transformation $(\phi |+\rangle_{\mathbf{e}_\sigma}, \chi|+\rangle_{\mathbf{e}_\sigma})^t \rightarrow (-\chi|-\rangle_{\mathbf{e}_\sigma}, \phi|-\rangle_{\mathbf{e}_\sigma})^t$, represented by operator 
\begin{equation}
    M = \begin{pmatrix}
        0 & -\sigma^z \\ \sigma^z & 0
    \end{pmatrix}.
\end{equation}
Applying this transformation to Hamiltonian (\ref{H complete 4x4}) yields 
\begin{equation}
    \tilde{H} = M^\dagger H M = \begin{pmatrix}
        \sigma^z(\hat{\bm{\gamma}}_+ - \Delta )\sigma^z  & -\sigma^z\hat{\bm{\gamma}}_-\sigma^z \\ 
				-\sigma^z\hat{\bm{\gamma}}_-\sigma^z & \sigma^z\left(\hat{\bm{\gamma}}_+ + \Delta \right)\sigma^z
    \end{pmatrix} .
    \label{eq:H-transformed}
\end{equation}
Comparing Hamiltonians (\ref{eq:H-transformed}) and (\ref{H complete 4x4}), we see that if 
\begin{equation}
    \{\hat{\bm{\gamma}}_+ , \sigma^z \} = 0, \hspace{0.5cm} \text{and} \hspace{0.5cm} \left[\hat{\bm{\gamma}}_-, \sigma^z \right] = 0 
\end{equation}
(that is, if only linear terms are kept in the momentum expansion of $\hat{\bm{\gamma}}_\pm$ in Eqs. (\ref{gammas})), then 
\begin{equation}
    \tilde{H} = - H, 
\end{equation}
which means exact particle-hole symmetry. Quadratic terms in the momentum expansion break this symmetry, 
producing an asymmetry of the order of $\Delta / \epsilon_F \ll 1$ 
relative to the otherwise particle-hole symmetric background.

\subsection{Solving the full Hamiltonian}
\label{subsec: BP Deep states solutions}

Now we turn to deep bound states in the intermediate range $R \sim \xi$ . In this preliminary section, we focus on a $Q=1$ skyrmion, and keep the presentation as general as possible. We consider Hamiltonian (\ref{H complete 4x4}) and use the expansion (\ref{gammas}), only keeping its dominant terms, which is valid for $R \sim \xi \gg a$. In this approximation, only differential operators acting on $y$ are kept, since the `slow' coordinate $x$ will only slightly affect the energy; thus we fix it at its mean value which is, by symmetry, $x = 0$. Technically, this implies setting $A_y^z = 0$ in $\hat{\bm{\gamma}}_-$, which is further supported by the fact that the dominant potential term is proportional to $\mathbf{e}_\sigma \cdot \bm{\sigma}$. With $\mathbf{e}_\sigma$ assumed to be orthogonal to $\mathbf{e}_z$, $A_y^z\sigma^z$ will only act in second order of perturbation theory. Stated differently, by analyzing the electron motion only along the `fast' coordinate $y$, we study the `gross' structure of the bound-state energy levels, temporarily ignoring the `fine' structure: the sublevels, appearing due to quantization of electron motion along the `slow' coordinate $x$. The approximations above imply 
\begin{equation}
	\begin{split}
	    \hat{\bm{\gamma}}_+ &= vA_y^\parallel \mathbf{e}_\sigma\cdot \bm{\sigma} \equiv -\gamma_+\mathbf{e}_\sigma\cdot \bm{\sigma},\\
	    \hat{\bm{\gamma}}_- &= v\hat{p}_y = \hat{\gamma}_-. \label{approx gamma +-}
	\end{split}
\end{equation} 
The Hamiltonian (\ref{H complete 4x4}) can now be applied to bispinor $(\phi_1,\psi_1, \psi_2,\phi_2)^t$ where the spinor components $\phi_i$ and $\psi_i$ are given in the spin eigen basis $|\pm\rangle_{\mathbf{e}_\sigma}$ along $\mathbf{e}_\sigma$.

The corresponding Schr\"odinger equation thus reads 
\begin{equation}
	\begin{split}
		&(1)\hspace{0.8cm} (\Delta - \gamma_+ -E)\phi_1 + \hat{\gamma}_- \psi_2 = 0,\\
		&(2)\hspace{0.8cm} (\Delta + \gamma_+ -E)\psi_1 + \hat{\gamma}_- \phi_2 = 0,\\
		&(3)\hspace{0.8cm} \hat{\gamma}_- \phi_1 - (\Delta +\gamma_++E)\psi_2 = 0,\\
		&(4)\hspace{0.8cm} \hat{\gamma}_-\psi_1 - ( \Delta - \gamma_+ +E) \phi_2 =0.
	\end{split}
\end{equation}
The equations (2) and (3) allow us to express $\psi_1$ and $\psi_2$ as 
\begin{equation}
	\begin{split}
	    \psi_1 &= -\frac{1}{\Delta + \gamma_+-E} \hat{\gamma}_-\phi_2, \\
	    \psi_2 &= \frac{1}{\Delta + \gamma_+ + E} \hat{\gamma}_-\phi_1.
	\end{split}
\end{equation}
Paying attention to non-commutativity of $\hat{\gamma}_-$ and $\gamma_+$, we find
\begin{equation}
	\left[\hat{\gamma}_-^2 -  \frac{\hat{\gamma}_-(\gamma_+)}{\Delta + \gamma_+ \pm E}\gamma_- + \Delta^2 - \left(\gamma_+ \pm E\right)^2\right] \phi = 0. \label{Eq générale phi}
\end{equation}
The particle-hole symmetry (\ref{eq: particle-hole symmetry}) is explicit here, with polarization inversion $|+\rangle_{\mathbf{e}_\sigma} \to |-\rangle_{\mathbf{e}_\sigma}$ corresponding to replacing $\phi_1$ by $\phi_2$, with sign $+$ in equation (\ref{Eq générale phi}) associated with $\phi_1$.

\begin{figure}[ht]
	\centering \includegraphics[width=\linewidth]{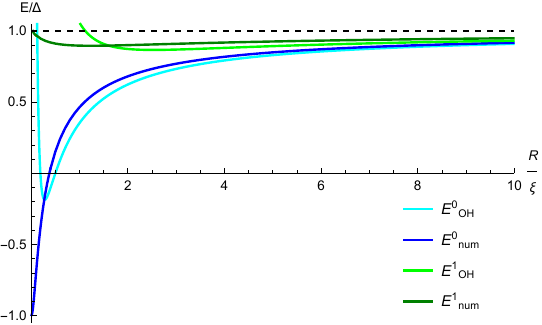} 
	\caption{Ground-state energy (with polarization $|+\rangle_{\mathbf{e}_\sigma}$) of Hamiltonian (\ref{H complete 4x4}) as a function of $\frac{R}{\xi}$, shown in units of $\Delta$. The numerical solution in blue is compared with the analytical one (\ref{Energy_OH_BP}) obtained with Harmonic approximation, plotted in cyan and valid for $\frac{R}{\xi} \gg 1$. The energy of the first excited state relative to the `fast' degree of freedom is represented in dark green, and again compared with its analytical counterpart (\ref{Energy_OH_BP}) in green. The abscissa of the crossing point between ground states energies $E_0^c$ and $E_0^v$ is $R^* \simeq 0.37\xi$.}
	\label{fig E BP}
\end{figure}

\begin{figure}[ht]
	\centering \includegraphics[width=\linewidth]{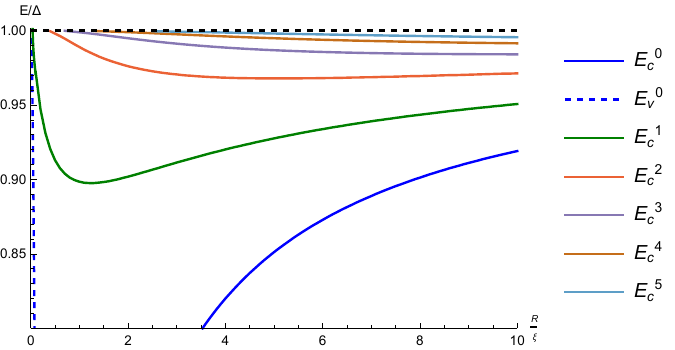}
	\caption{Excited-state energies $E_c^i$ (in $\Delta$ units) as functions of $\frac{R}{\xi}$, obtained by numerical solution of equation (\ref{Eq générale phi}). Indices $i$ represent the number of the wave-function zeros along $y$. The shown states are arising from the conduction band, and have symmetric counterparts, generated by the valence band. }
	\label{fig En(R) BP}
\end{figure}

This one-dimensional second-order differential equation can be easily solved numerically. Searching for energies $E$, for which the wave function is physical, leads to a discrete energy spectrum $E_n(R)$. These energies are depicted as functions of the BP skyrmion radius $R$ in Figs. \ref{fig E BP} and \ref{fig En(R) BP}. Figure \ref{fig E BP} shows numerical solutions for the ground-state energy and the energy of the first excited state (with respect to the `fast' degree of freedom $y$) in comparison with the analytical result (\ref{Energy_OH_BP}) obtained in the large-radius limit $R\gg\xi$. The agreement is excellent for $R \gtrsim 8\xi$, as expected. Note that the ground state is the only state whose energy crosses zero. 

Fig. \ref{fig En(R) BP} shows numerical results for the first excited states. Their structure is non-trivial for $R$ comparable to $\xi$, where the frequencies $E_c^{i+1} - E_c^{i}$ are far from constant. The harmonic description holds only for low-lying excited states, and for $R$ large against the coherence length $\xi$.

Note that this solution describes only the $y$-dependence of the wave function. As we already pointed out, it encapsulates the `gross' structure of bound-state spectrum. Each state of energy $E_n(R)$ is accompanied by a set of excited states separated by frequencies of the order of  $\hbar\omega_x$. This is due to the fact that for the `slow' degree of freedom $x$, the kinetic term remains small compared with main potential term even in the range $R \lesssim \xi$. In the picture above, the energy $E_c^{0}$ of the lowest conduction-band bound state crosses zero at an $R \sim \xi$. Combined with the particle-hole symmetry, this implies level crossing with bound states arising from the valence band. We will 
now see if this level crossing is avoided.

\subsection{Perturbation-theory analysis of crossing points}
\label{subsec : Perturbation theory}

Now we will treat the previously neglected terms in $\hat{\bm{\gamma}}_\pm$ of Eq. (\ref{gammas}) as perturbation of the dominant terms (\ref{approx gamma +-}). Figure \ref{fig En(R) BP} suggests that the only states allowed to intersect are those whose wave function has no zeros along the $y$ direction. Focusing on these states, we denote them as
\begin{equation}
	\Psi_i^c = \begin{pmatrix}
		\phi_i \\ 0 \\ \psi_i \\ 0
	\end{pmatrix}, \hspace{1cm}
	\Psi_i^v = \begin{pmatrix}
		0 \\ -\psi_i \\0 \\ \phi_i
	\end{pmatrix}, \label{états propres dégénérés pt de croisement}
\end{equation}  
where indices $i$ stand for the number of wave-function zeros along $x$. In first approximation, these states can be presented as $\phi_i(x,y) = \phi(y)\langle x|i\rangle$ and $\psi_i(x,y) = \psi(y)\langle x|i\rangle$, with $|i\rangle$ the one-dimensional harmonic oscillator eigenstates, and $\phi (y)$ the ground-state eigenfunction of Eq. (\ref{Eq générale phi}). Here $\psi = \frac{\hat{\gamma}^0_-}{\Delta + \gamma^0_+ } \phi$, with $\hat{\gamma}_\pm^0$ representing the zero-order expansion (\ref{approx gamma +-}) of $\left\|\hat{\bm{\gamma}}_\pm \right\|$.

An avoided crossing can only appear due to off-diagonal terms in the spin basis defined by $\mathbf{e}_\sigma$. In $\hat{\bm{\gamma}}_-$, among such terms the dominant one is $v A_y^z \sigma^y$ \footnote{We make a change of basis that amounts to a permutation of spin space indices $(x,y,z)$, replacing $(x,y,z)$ by $(z,x,y)$.}. This will split the crossing energies 
\begin{equation}
    E_i^0 = \langle \Psi_i^c | H_0 | \Psi_i^c \rangle = \langle \Psi_i^v | H_0 | \Psi_i^v \rangle 
\end{equation}
with $H_0$ made of $\hat{\gamma}_\pm^0$, by the amount \cite{Landau1981Quantum} 
\begin{equation}
	\delta_{ij} = \langle \Psi_i^c |V| \Psi_j^v\rangle , \hspace{0.7cm} V = vA_y^z \sigma^y \otimes \sigma_x,
\end{equation}
with $\sigma_x$ coupling spinors. To estimate the splitting, we approximate $A_y^z \sim \Delta \frac{\xi}{R}\frac{\tilde{x}}{1+\tilde{y}^2}$, and observe that the integral
\begin{equation}
    \int_{-\infty}^{+\infty} d\tilde{y}\, \frac{|\phi|^2 + |\psi|^2}{1+\tilde{y}^2} 
\end{equation}
appears to be numerically near unity for $R \sim \xi$. Using standard properties of harmonic oscillator eigenstates, we find 
\begin{equation}
	\delta_{ij} \sim \Delta \frac{\xi}{R} \langle i | \hat{\tilde{x}} |j\rangle \sim \Delta \frac{\xi}{R} \left(\epsilon\frac{\xi}{R}\right)^{1/4} \left(\delta_{i,j+1} + \delta_{i+1,j}\right). \label{expression gap}
\end{equation}
In reality, the wave functions of crossing states are more complicated, and $A_y^z$ is not a separable function. However, since $A_y^z$ is an odd function of $x$, it couples states of different parities, meaning $\delta_{ij}$ is expected to be non-zero if $|i-j|$ is an odd integer, and to realize its maximum for $|i-j|=1$. 

The other off-diagonal term is proportional to $A_x^y$, an odd function of $y$, and is irrelevant since there is no crossing between eigenstates of different parity along $y$, see Fig. \ref{fig En(R) BP}.

These arguments suggest that, to second order in momentum expansion of Eq. (\ref{gammas}), there is no avoided crossing between levels of the same spatial symmetry. Looking at the ratio 
\begin{equation}
    \frac{\delta_{01}}{\hbar\omega_x} \simeq \frac{1}{2} \left[ \epsilon \frac{\xi}{R} \left(1 + \frac{\xi}{2R}\right)^3 \right]^{-1/4} ,
\end{equation}
we see that it is larger than one as long as the skyrmion radius is larger than 
\begin{equation}
	R_c \sim \epsilon^{1/4} \xi \ll \xi 
	\label{eq : critical radius harmonic structure}
\end{equation}
with $\epsilon = \frac{m_y}{m_x}\frac{a}{\xi} \lll 1$.
This appears to be the radius for which the  characteristic frequency of the `slow' degree of freedom 
\begin{equation}
	\hbar w_x = \Delta \left(\frac{\xi}{R}\right)^{3/2} \sqrt{2\epsilon}\sqrt{1+\frac{\xi}{2R}} \label{ecart niveau energie x}
\end{equation}
becomes of the order of $\Delta$. Thus, as long the present discussion makes sense, the gaps opening at avoided crossing points are much larger that the characteristic spacing between the energy levels. This observation implies that the off-diagonal coupling term $A_y^z\sigma^z$ will avoid all possible crossings between levels of the same spatial symmetry along the $x$ axis. Since the number of crossing levels is large, the precise energy structure is difficult to identify. 
However, the qualitative picture of the bound-state evolution as a function of the skyrmion radius $R$ 
is afforded by the textbook argument for electron terms in a diatomic molecule \cite{Landau1981Quantum} 
and is shown in Fig. \ref{fig: Structure interne bande}. 
The argument states that, for a single-parameter Hamiltonian, two electron terms of the same symmetry cannot cross. 
In the present problem, the single parameter is the skyrmion radius $R$, and the avoided crossing is ensured by 
the term $V = vA_y^z \sigma^y \otimes \sigma_x$ \footnote{This term only couples 
states of opposite polarization and different spatial symmetry along $x$. Applying it as a perturbation 
does however produce states of superposed polarizations, meaning they will interact 
with states of any spatial symmetry due to other terms in the Hamiltonian that have been previously ignored. 
} 

\subsection{Physics picture}
\label{subsec: BP physics picture}

\begin{figure}[ht]
	\centering \includegraphics[width = \linewidth]{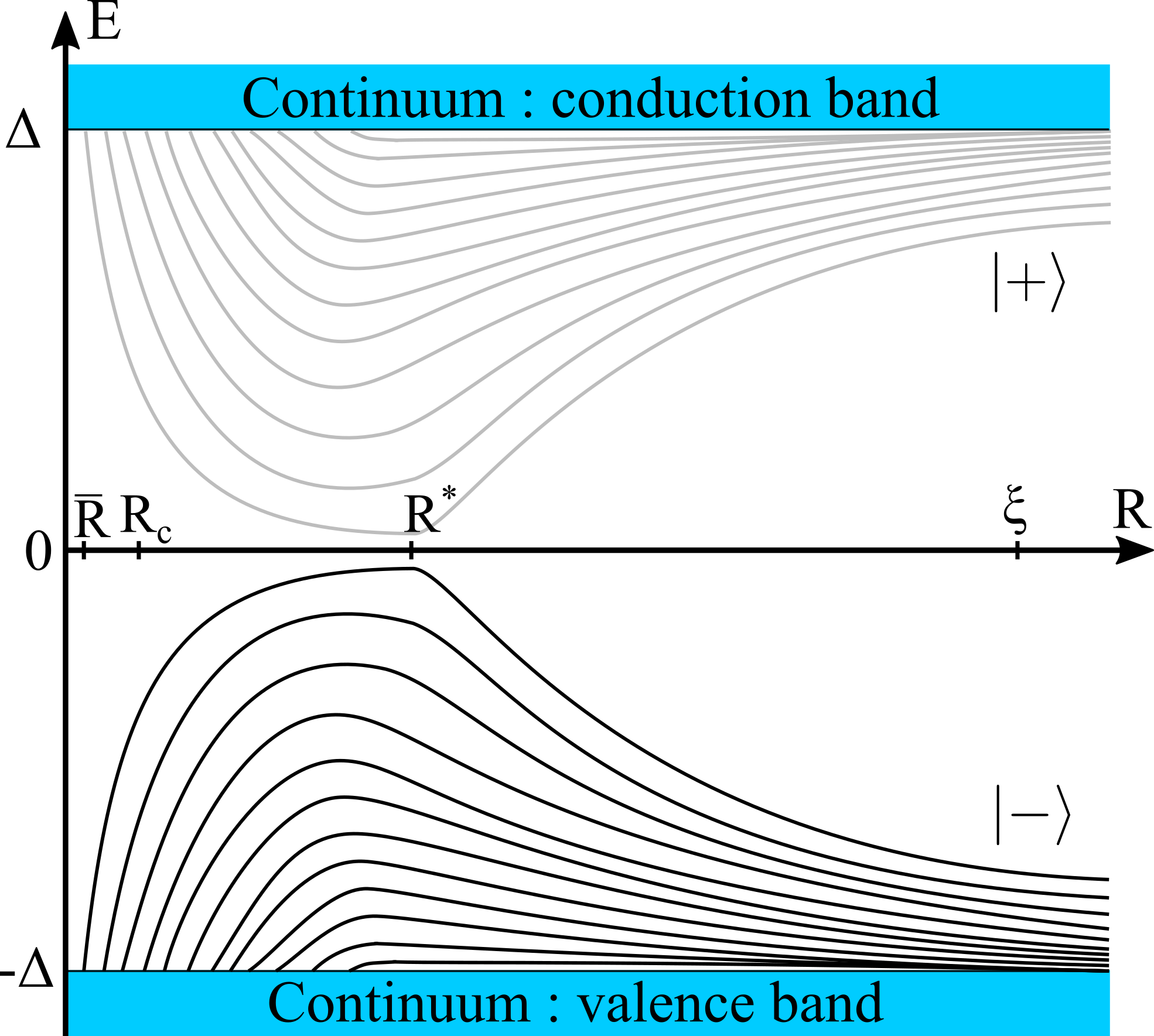}
	\caption{ Evolution of the bound-state levels as a function of the BP skyrmion radius $R$, key features. $R^*$ represents the largest radius
            allowing for level crossing before perturbation, see Fig. \ref{fig E BP}. And $R_c$ is the radius for which the ‘slow’ frequency $\omega_x$ becomes  comparable with $\Delta$, see Eq.~(\ref{eq : critical radius harmonic structure}). }
	\label{fig: Structure interne bande}
\end{figure}

The analysis above allows us to draw the following physics picture. For large BP skyrmion radia $R \gg \xi$, the bound states remain shallow: that is, they lie near the edge of the band they originated from: $| \varepsilon | \ll \Delta$. For both $x$ and $y$ coordinates, the low-lying states are close to those of a harmonic oscillator, see Eq. (\ref{Energy_OH_BP}). As $R$ decreases to approach the coherence length $\xi$, these low-lying states become more and more profound, and approach the gap center at $R \sim R^*$. At lower radia $R \lesssim R^*$, the energy gap is filled by a ladder of bound states of nearly harmonic structure along $x$, with a characteristic level spacing $\hbar \omega_x$. As $R$ decreases further, the harmonic structure along $x$ is lost, and levels split to reach an energy difference of about $\Delta$ at $R \sim R_c$, before the last bound state disappears at $R = \bar{R}$, with $a \ll \bar{R} \sim \sqrt{\frac{m_y}{m_x}} \sqrt{a \xi} \ll \xi$. For $R < \bar{R}$, there are no bound states any more, a result that holds as long as the continuum approximation is valid, that is for $R\gg a$. Evolution of bound states with $R$ is sketched in Fig. \ref{fig: Structure interne bande}.

Note that, generally, the eigenstates of Hamiltonian (\ref{H complete 4x4}) are spin-polarized non-uniformly. In sections~\ref{sec: BP Profile, small energies} and \ref{sec: BP Deep bound states} above, the Hamiltonian and the resulting eigenstates  
were expressed in the basis obtained via transformations  $P$ and $U(\mathbf{r})$. The physical wave function must then be reconstructed from the found solution $\Psi$ via
\begin{equation}
    \Psi_{\text{physical}}(\mathbf{r}) = U(\mathbf{r}) P \Psi(\mathbf{r}).
\end{equation}
Spin rotation $U(\mathbf{r})$ is non-uniform. Thus, generally, so is the spin polarization of the wave function. 
However, for a large BP skyrmion ($R \gg \xi$), the wave functions of low-lying states are localized well within the skyrmion core ($r \ll R$), 
where the N\'eel order parameter points along $\hat{z}$, and thus $U(\mathbf{r})$ is close to $\pi$-rotation around $\hat{z}$. 
Therefore, for $R \gg \xi$, the physical wave function is spin-polarized opposite to $\hat{x}$, as shown 
in Figs. \ref{fig: wave function BP} and \ref{fig: wavefunction DW}. 
However, for a small skyrmion ($R \lesssim \xi$), spin polarization of the wave function 
becomes substantially non-uniform even for low-lying bound states \cite{Davier_2023}.\\

Being parameterized by a single length scale and having a simple and elegant analytic profile, Belavin-Polyakov 
skyrmion is a tractable example \textit{par excellence}, allowing one to analyze skyrmion-electron bound 
states. Moreover, weak single-ion anisotropy affects the skyrmion shape only far away from the center ($r \gg R$) 
\cite{Komineas_2020, Voronov_1983}, and thus does not influence the low-lying bound states whose wave functions are localized 
well within the skyrmion core. However, the BP profile is relevant in a parameter range where the skyrmion lifetime 
appears to be short \cite{AF_Sk_Bessarab_2019}.  
Skyrmion-electron bound states only increase the skyrmion energy, and thus reduce its stability even further. 
We now turn to a different profile, which is described by \textit{two} length scales rather than one: 
that of a `domain wall' (DW) skyrmion. The DW profile becomes relevant near the transition to the modulated 
phase \cite{Bogdanov_1989}, where skyrmions tend to have longer lifetimes \cite{AF_Sk_Bessarab_2019}.

\section{Domain wall profile}
\label{sec: Domain wall profile}

\begin{figure}[ht]
	\centering   \includegraphics[width = \linewidth]{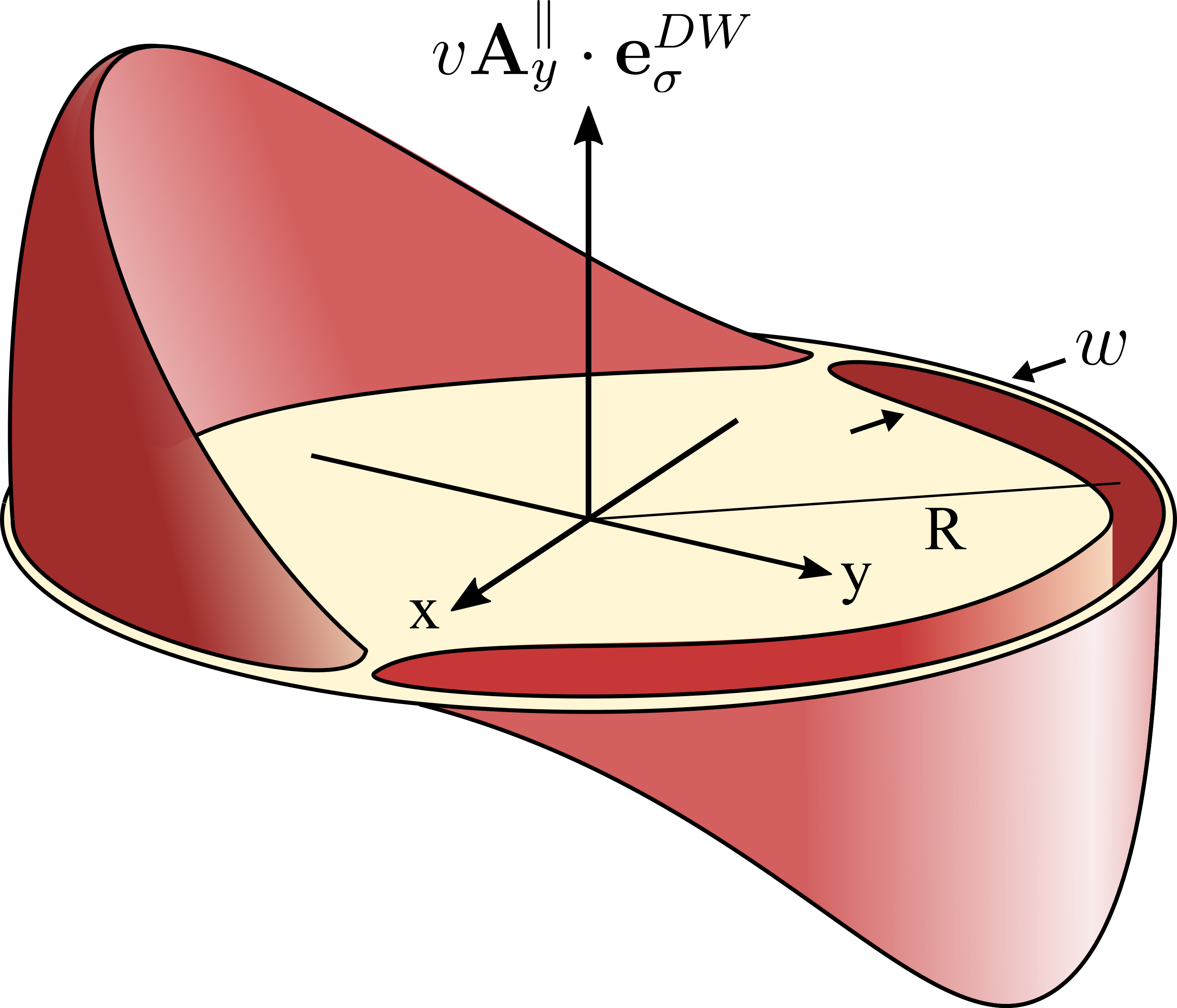} 
	\caption{A sketch of the dominant potential term $v\mathbf{A}_y^{\parallel}\cdot\mathbf{e}_\sigma^{DW}$, for a domain wall profile of radius $R$, large compared with the wall width $w$.}
	\label{fig: Potentiel_DW}
\end{figure}

In this section, we study the so-called Domain Wall (DW) skyrmion profile defined by 
\begin{equation}
	\theta(r) = \pi - 2 \arctan \left(\frac{\sinh(R/w)}{\sinh(r/w)}\right) \label{def profil DW},
\end{equation}
with the skyrmion radius $R$ and the domain wall width $w$. The name `domain wall skyrmion' points to a nearly constant $\theta (r) \ll 1$ within the skyrmion radius $R$, undergoing a smooth variation within the distance $w \ll R$ around $r = R$, to become nearly constant ($\theta \approx \pi$) again at $r > R$. That is, such a skyrmion can be seen as a circular domain wall of radius $R$ and width $w$. We consider a DW skyrmion with winding number $Q=1$ and an arbitrary helicity $\gamma$. In the relevant limit $R \gg w$ \cite{komineas2020Large_radius}, we can make an approximation 
\begin{equation}
	\theta'(r) \simeq \frac{1}{w\cosh\left(\frac{R-r}{w}\right)} \simeq \frac{\sin \theta}{w}, \label{O' sim sinO} 
\end{equation} 
which implies, near $r = R$, that 
\begin{equation}
	\frac{\sin\theta}{r} \sim \frac{\sin\theta}{R} \ll \frac{\sin\theta}{w} \simeq \theta'.
\end{equation}
The inequality above allows us to simplify the expression for the gauge field in Eq. (\ref{Expression_explicit_A_y^x}). For the dominant term, this leads to \footnote{This is equivalent to the inequality $\partial_i\theta \gg \partial_i \phi$, which allows us to keep only the first term in the expression (\ref{definition A_i}) for $\mathbf{A}_i$.}
\begin{equation}
	v\mathbf{A}_y^{\parallel} \cdot \bm{\sigma} \simeq -\Delta \frac{\xi}{2w}\sin\theta \sin\alpha \, \mathbf{e}_\sigma^{DW} \cdot\bm{\sigma} \label{vA para DW}
\end{equation}
with $\mathbf{e}_\sigma^{DW} = (\sin(\alpha + \gamma),- \cos (\alpha + \gamma) ,0)^t$. Helicity then plays the same role as for the BP profile: it only defines the polarisation of the bound states. This potential has a non-trivial shape, depicted in Fig. \ref{fig: Potentiel_DW}. It has two minima of depth $\Delta \frac{\xi}{2w}$ located at points $\pm \mathbf{r}_0 = \pm R\mathbf{e}_y$ with opposite  polarization relative to $\mathbf{e}_\sigma^{DW}$.

\subsection{Shallow bound states}
We start by treating the limit $w \gg \xi$, where the potential is small against $\Delta$. Direct inspection shows that, near the minima  $\pm \mathbf{r}_0$, the dominant repulsive term can be estimated as 
\begin{equation}
	\begin{split}
		\frac{\left(A_y^z\right)^2}{2m_y^*}  \sim \Delta \frac{\xi^2}{8R^2}x^2 
   \ll vA_y^\| \sim \Delta \frac{\xi}{2w} .
	\end{split}
\end{equation} 
That is, repulsion remains small compared with the dominant attractive potential $v\mathbf{A}_y^\parallel\cdot\bm{\sigma}$, and can be neglected altogether. Cross-terms $\{\hat{p}_i,A_i^z\}\sigma^z$ are also small and off-diagonal in the basis of $\mathbf{e}_\sigma^{DW}$. Thus, in the first approximation, they can also be ignored. 

The potential can be diagonalized with the help of an operator 
\begin{equation}
	P(\alpha) = \frac{1}{\sqrt{2}}\left(1-i\mathbf{e}_r\cdot \bm{\sigma}\right) 
\end{equation}  
with $\mathbf{e}_r = \left(\cos(\gamma + \alpha), \sin (\gamma + \alpha), 0\right)^t$, which produces additional terms when 
sandwiching the operator $\hat{p}_i^2 \rightarrow P^\dagger(\alpha) \hat{p}_i^2 P(\alpha)$ 
\footnote{This modifies the Peierls substitution in Eq. (\ref{Peierls}). }. Nevertheless, depending only on the polar angle $\alpha$, operator $P$ varies on the characteristic scale $R$, and thus produces only terms of the order $\Delta \frac{\xi^2}{R^2}$ (and even smaller for $\hat{p}_x$), negligible compared with $vA_y^\parallel$.

The resulting Schrödinger equation for bound states near the point $+ \mathbf{r}_0$ is then 
\begin{equation}
	\begin{split}
		\left[-\frac{\xi^2}{2w^2}\frac{\partial^2}{\partial \tilde{y}^2} - \frac{\epsilon}{2}\frac{\xi^2}{R^2} \frac{\partial^2}{\partial \tilde{x}^2} - \frac{\xi}{2w} \sin\theta \sin\alpha \right] \psi^{\uparrow } = \frac{\varepsilon}{\Delta}\psi^{\uparrow }
	\end{split}
\end{equation}
with $\tilde{x} = \frac{x}{R}$ and $\tilde{y}=\frac{y-R}{w}$, and the spin projection $\uparrow$ along $\mathbf{e}_\sigma^{DW}$. Note that here the anisotropy is much stronger than in the BP case: on top of the effective-mass anisotropy, the potential itself is now strongly anisotropic. Such a strong anisotropy again invites the Born-Oppenheimer approximation. The low-energy limit $w \gg \xi$ is also appropriate for a harmonic expansion, since the kinetic term is small compared with the depth of the potential well. The potential can be expanded locally as 
\begin{equation}
	-\sin(\alpha) \, \text{sech}\frac{r-R}{w} \simeq -1 + \frac{\tilde{x}^2 + \tilde{y}^2}{2} 
\end{equation}
yielding the energy levels 
\begin{equation}
	\frac{\varepsilon_m^n}{\Delta} = -\frac{\xi}{2w}\left[1 - \left( n + \frac{1}{2} \right)\sqrt{\frac{2\xi}{w}}\right] + \left( m + \frac{1}{2} \right) \frac{\xi}{R} \sqrt{\frac{\epsilon}{2}\frac{\xi}{w}}. \label{Enm DW}
\end{equation}

\begin{figure}
    \centering
    \includegraphics[width=\linewidth]{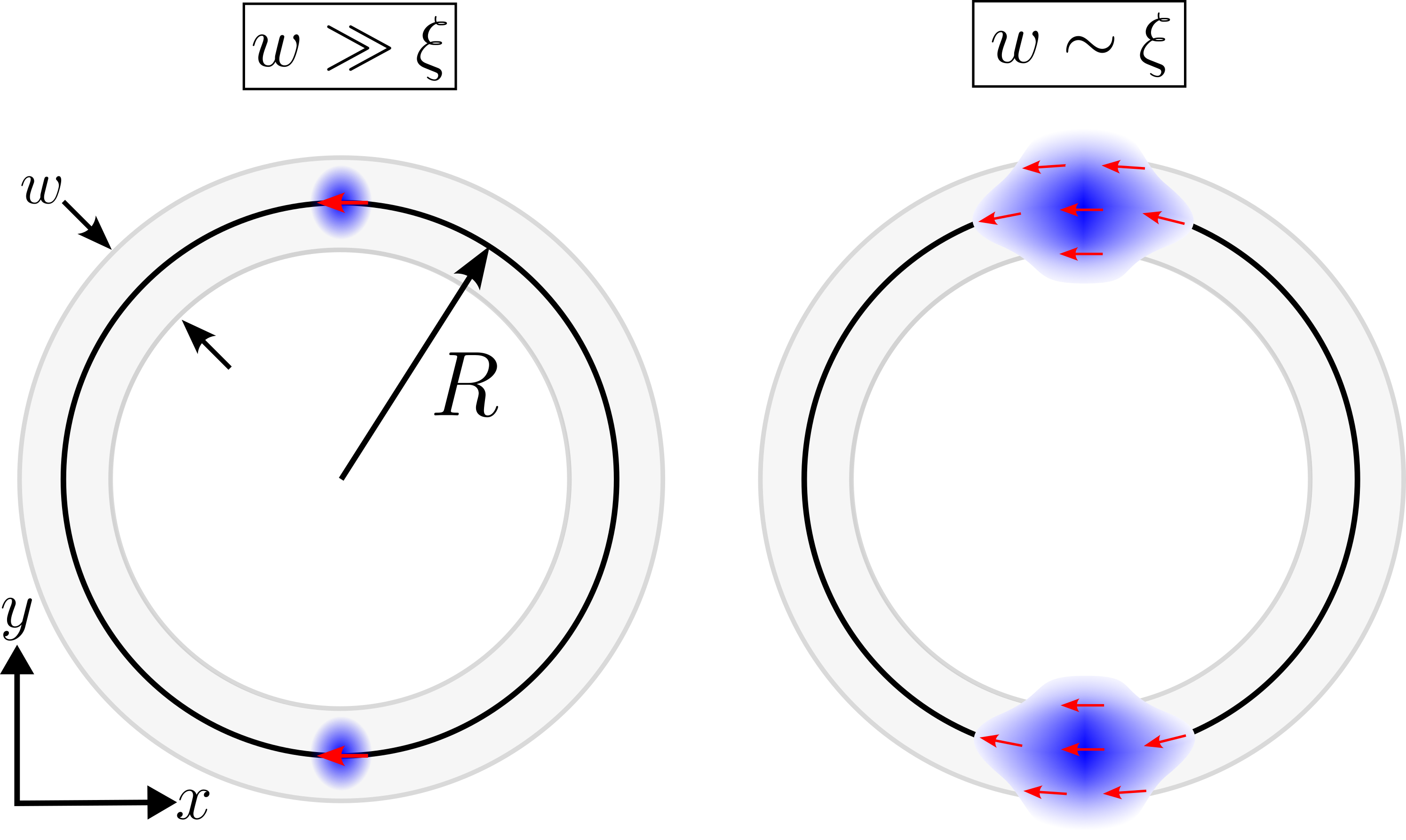}
    \caption{Density plot of the ground-state wave function (see Eq. (\ref{Enm DW})) of an electron, bound to a domain-wall skyrmion in limits $w \gg \xi$ and $w\sim \xi$. For a domain-wall skyrmion with a large wall width $w \gg \xi$, lowest electron bound states lie deep in the effective potential well formed by the domain wall. Their wave functions are thus close to those of a substantially anisotropic two-dimensional harmonic oscillator, which implies strong localization of the trapped electron inside the domain wall, as depicted in the left panel of the figure. With the wall width $w$ decreasing to approach $w\sim \xi$, the effective potential well becomes more and more shallow, which makes the wave function of the low-lying electron states spill outside the domain wall, as depicted in the right panel above. In both cases, the states localized near the bottom well at $(0, \pm R)$ are polarized along $\pm \mathbf{e}_\sigma^{DW}$, pointing along $- \mathbf{e}_x$ at the bottom well. 
    Away from $(0, \pm R)$, local spin polarization of the bound state 
    remains tangential to the domain wall, as shown in the right panel.
    The characteristic spread of the ground state wave function along $y$ is $\bar{y} \sim w \left(\frac{\xi}{w}\right)^{1/4}$, 
    while along $x$ it is $\bar{x} \sim R \left(\epsilon \frac{w\xi}{R^2}\right)^{1/4}$. Thus, by contrast with $\bar{y}$, 
    the bound-state range $\bar{x}$ along the domain wall is sensitive not only to the domain wall width $w$, 
    but also to the skyrmion radius $R$.
    }
    \label{fig: wavefunction DW}
\end{figure}

The spectrum above is similar to that of shallow bound states for the BP profile in Eq. (\ref{Energy_OH_BP}), 
except for an overall factor $\frac{1}{2}$: as a function of $w$, the potential well is a half of what it was for the BP profile as a function of $R$. The frequency of the `slow' degree of freedom is however smaller, due to the additional anisotropy of the potential. The wave functions of these  states are those of a two-dimensional anisotropic harmonic oscillator. The key difference relative to the BP skyrmion is that, here, the electron is predominantly localized at the two opposites edges of the DW skyrmion rather than near its center, compare Figs. 
\ref{fig: wave function BP} and \ref{fig: wavefunction DW}.

Just as for the BP skyrmion, this low-energy description breaks down with $w$ decreasing to become comparable with $\xi$: at this point, the full Hamiltonian (\ref{H complete 4x4}) has to be solved. 

\subsection{Deep bound states}
\begin{figure}[ht]
	\centering \includegraphics[width=\linewidth]{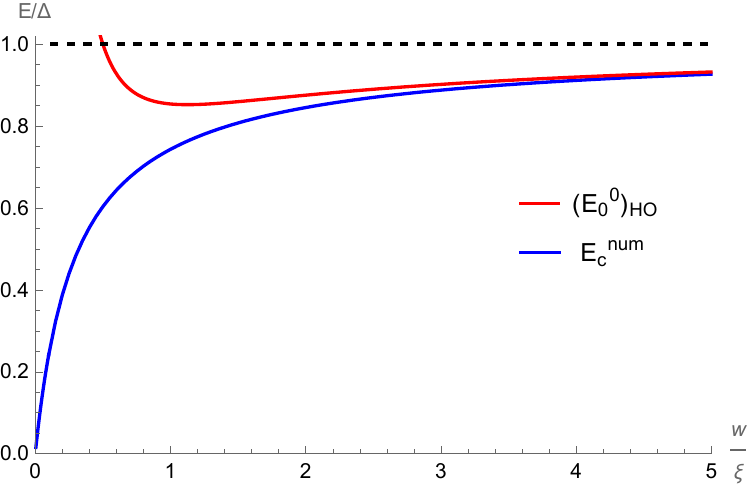} 
	\caption{Ground state energy (in units of $\Delta$) given as a function of the domain wall width $w$ in units of $\xi$. The numerical solution (in blue) is compared with the analytical result (\ref{Enm DW}), in red, obtained for $w\gg \xi$. The agreement is excellent for $w\gtrsim 4\xi$, as indeed it should be.}
	\label{fig:Tracé E_0(w) DWB}
\end{figure}

\begin{figure}[ht]
	\centering \includegraphics[width=\linewidth]{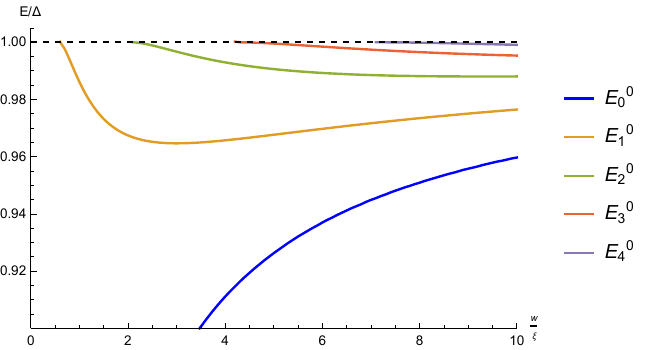}
	\caption{Numerical results obtained from the resolution of equation (\ref{Eq générale phi}) with DW profile. The energies $E_n^0(R)$ (in units of $\Delta$) are given as a function of 
 wall with $w$ (in units of $\xi$). Subscript $n$ stands for the number of wave-function zeros along $y$. Superscript 0 indicates that corresponding states have no zeros along the $x$ coordinate.}
	\label{fig: En(R) DW}
\end{figure}

Equation (\ref{Eq générale phi}) that we derived to treat the BP profile becomes even more accurate for the DW configuration, since the ratio $|A_y^z|/|A_y^{\|}|$ is now smaller than for BP skyrmion. Again, we solve Eq. (\ref{Eq générale phi}) numerically, and find the `gross' spectrum $E_n^0$ with $n$ the number of wave-function zeros along $y$. The results are shown in Fig. \ref{fig:Tracé E_0(w) DWB} for the ground state, and Fig. \ref{fig: En(R) DW} for the excited states.

Relative to the BP profile, the key difference can be observed already for the ground state:
it never crosses zero energy, even in the limit $w \to a$, meaning there is no crossing between states generated by the valence and conduction bands. 
Thus, the in-gap structure will be much simpler here than for a BP skyrmion. The difference arises from the factor $\frac{1}{2}$ in front of the square brackets in Eq. (\ref{Enm DW}) as compared with Eq. (\ref{Energy_OH_BP}): for a DW skyrmion with a given $w$, the attractive potential is twice weaker than for an equal-radius ($R = w$) BP skyrmion. The DW excited-state structure is very similar to the BP case, even if the bound states are about twice as shallow.

The harmonic structure for sub-levels associated with the `slow' degree of freedom $x$ holds longer for the DW profile. Indeed, this time 
\begin{equation}
	\frac{\hbar\omega_x}{\Delta} = \frac{\xi}{R}\sqrt{\frac{\epsilon}{2}\frac{\xi}{w}} \label{h wx DW}
\end{equation} 
becomes comparable with unity for
\begin{equation}
	w_c = \frac{m_y}{m_x} \frac{\xi^2}{2R^2} a. \label{w_c}
\end{equation}
Thus, for $R \gtrsim \sqrt{\frac{m_y}{m_x}}\xi$, the energy $\hbar\omega_x$ remains always small compared with the gap $\Delta$. Which means that, in the continuum approximation, harmonic structure persists for any radius $R$. \\

As in the BP case, the dominant repulsive term $\frac{\left(A_y^z\right)^2}{2m_y^*}$ could be responsible for the loss of bound states at a lower wall width $\bar{w}$. However, since harmonic description holds for the slow degree of freedom, this term can be estimated as 
\begin{equation}
	\left\langle \frac{\left(A_y^z\right)^2}{2m_y^*}\right\rangle \sim \Delta \frac{\xi^2}{R^2} \langle\tilde{x}^2\rangle  \sim \Delta \frac{\xi^3}{R^3} \sqrt{\epsilon\frac{w}{\xi}}.
\end{equation}
Therefore, this repulsion can never compensate the attraction $vA_y^\parallel \sim \Delta \frac{\xi}{w}$. Which means that, for a DW profile with $w \ll R$, bound states are always present, at least in the continuum description.\\ 

\begin{figure}[ht]
	\centering \includegraphics[width=8cm]{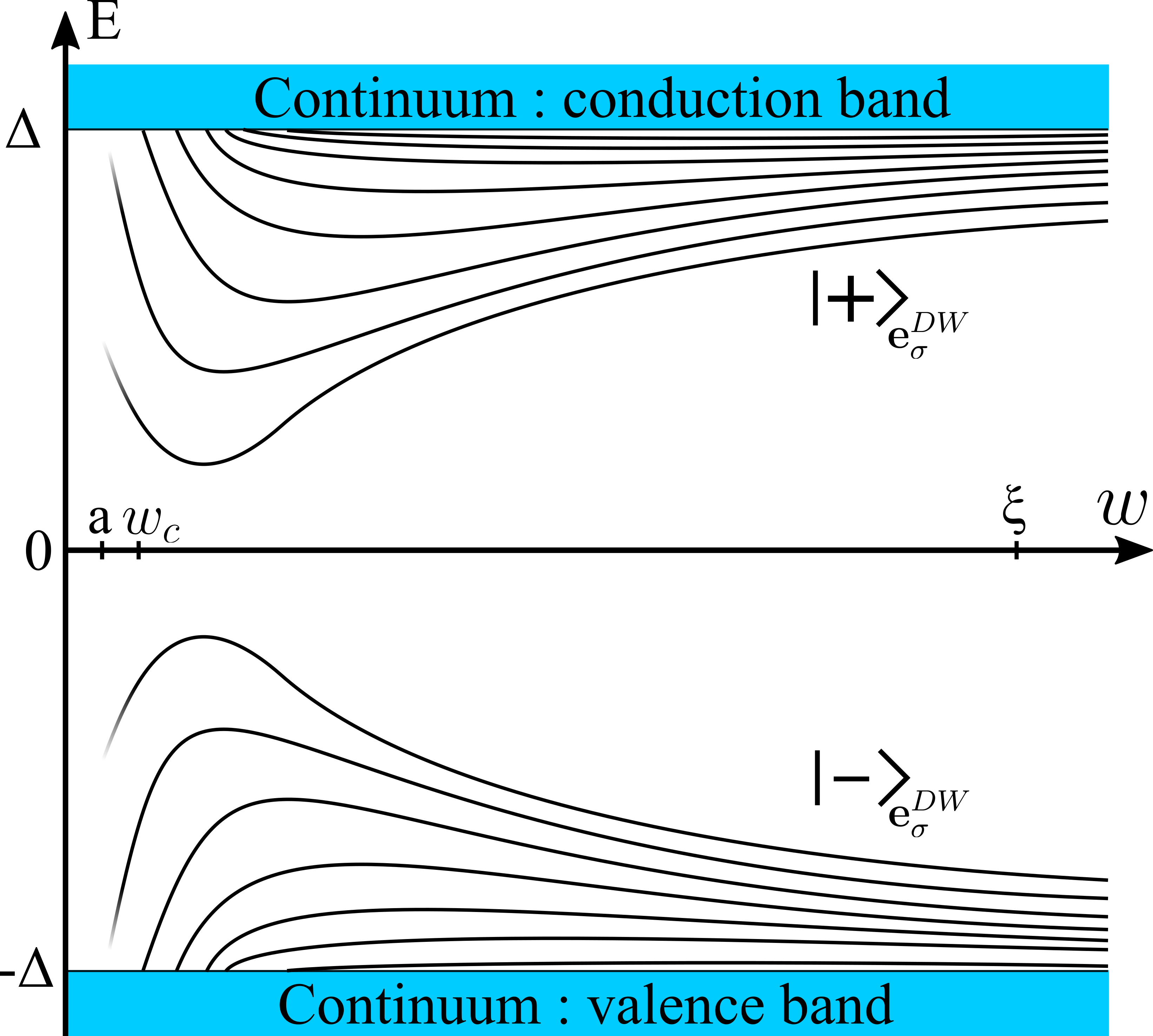}
	\caption{The in-gap structure for a DW profile. There is no crossing between states arising from valence- and conduction bands. The numerical solutions of equation (\ref{Eq générale phi}) hold until $w$ becomes comparable with critical width $w_c$, where the fine-structure level spacing becomes comparable with $\Delta$, see Eq.~(\ref{w_c}). 
 }
	\label{fig: In-gap structure DW}
\end{figure}

For the DW profile, evolution of the bound-state spectrum as a function of $w$ looks similar to what one sees for the BP skyrmion as a function of $R$: compare Figs. \ref{fig: In-gap structure DW} and \ref{fig: Structure interne bande}. At the same time, we observe two important differences: (i) there is no critical DW width $\bar{w}$ below which bound states disappear -- at least, as long as $w \ll R$. And, since there is no avoided crossing, (ii) polarisation of bound states remains pure, that is  $|+\rangle_{\mathbf{e}_\sigma}$ or $|-\rangle_{\mathbf{e}_\sigma}$. Note that, in the leading approximation, the skyrmion radius $R$ has no impact on the `gross' structure of bound states: see the first term in Eq. (\ref{Enm DW}). However, $R$ does affect the `fine' structure of bound sates, see $\hbar \omega_x$ in Eq. (\ref{h wx DW}).

Having gained insight into the bound-state spectrum for both the BP and the DW profiles, 
we now turn to the contribution of bound states to the skyrmion energy at half-filling.

\section{Energetics at half-filling}
\label{sec: Half filling energetics}

We consider a material with one electron per site, 
at zero temperature. In the N\'eel phase, all the 
negative-energy states are occupied, while all the positive-energy states are empty, 
see Figs. \ref{fig: Structure interne bande} and \ref{fig: In-gap structure DW}. 
Filled bound states, generated by the valence band, rise above $E = -\Delta$ (the top of the valence band in the absence of the skyrmion), thus making a positive contribution to the skyrmion energy. 

The electron contribution to the skyrmion energy may be defined as the difference 
between the sums of occupied-state energies with and without the skyrmion
\begin{equation*}
	E_T = \sum_i \underbrace{(E^v_i)^{\text{Bloch}} + (E_i^v)^\text{bound}}_{\text{with\;skyrmion}} - \sum_i\underbrace{(E^v_i)^{\text{Bloch}}}_{\text{without\;skyrmion}}.
 \label{eq:E_T}
\end{equation*}
Calculation of $E_T$ requires understanding the effect of the skyrmion on itinerant electron states. Let $N_\text{bs}(R)$ be the microscopic number of valence-band bound states with polarization $|-\rangle_{\mathbf{e}_\sigma}$, for a single valley $\Sigma_i$ (see fig \ref{fig: MBZ}). 
On the other hand, in Sec. \ref{sec: Hamiltonian construction} we saw that in the uniform N\'eel state all the electron eigenstates are doubly degenerate. Thus there are $N_\text{bs}$ states with the same unperturbed energy, but polarization $|+\rangle_{\mathbf{e}_\sigma}$, that are repelled from the skyrmion, resulting in a smaller density near the skyrmion center, and a smaller energy. The intra-band energy shift for one of these states can be estimated as 
\begin{equation}
	\delta \varepsilon_i = v \int d^2r A_y^\parallel |u_i|^2 \sim v \int d^2r A_y^\parallel |u_0|^2
\end{equation}
with $u_i(\mathbf{r})$ the $i^{\text{th}}$ Bloch function. Since the relevant states are close to the valence band maximum, we make the approximation $u_i \sim u_0 \sim \frac{1}{L}$ (with $L$ the characteristic system size) in the spirit of the rigid band theorem \cite{Kittel1963}. This yields the energy shift
\begin{equation}
	\delta \varepsilon \simeq \Delta \frac{\xi}{ R}\frac{R^2}{L^2}\int_0^{L/R} \frac{z}{1+z^2}dz \simeq \Delta \frac{\xi}{ R}\frac{R^2}{L^2} \ln \frac{L}{R}, 
 \label{eq:shift}
\end{equation}
which is a factor of $\frac{R^2}{L^2} \ln \frac{L}{R}$ smaller than the characteristic energy $\Delta \frac{\xi}{R}$ of the bound states. 
In the thermodynamic limit $L \ggg R$, the characteristic energy of these $N_\text{bs}$ $|+\rangle_{\mathbf{e}_\sigma}$-polarized states is thus negligible relative to the characteristic energy $\Delta \frac{\xi}{R}$ of the bound states. The impact of the skyrmion on the energy of Bloch states is therefore tiny compared with the effect of bound state formation, since there is an equal number $N_\text{bs}$ of states of each polarization. The $N_\text{bs}$  $|+\rangle_{\mathbf{e}_\sigma}$-polarized  states can then be ignored when calculating the electron contribution $E_T$ to the skyrmion energy.

Finally, consider the other states of the same valley, that are described by the effective-mass Hamiltonian, but lie sufficiently deep in the band, so that the corresponding $|-\rangle_{\mathbf{e}_\sigma}$-polarized states do not become bound. These states will experience an energy shift similar to (\ref{eq:shift}) with a sign, dependent on their polarization. Since states of opposite polarization are initially degenerate, the energy shifts of a pair of states will tend to compensate. The resulting statement is that the total energy $E_T$ can be approximated by the sum over occupied bound-state energies only. \\

Before estimating $E_T$, a remark concerning electron-electron interaction. While its careful treatment is beyond the scope of this work, for a sufficiently large skyrmion electric neutrality will enforce uniform electron density: in each valley, one spin polarization is attracted to the skyrmion center to form bound states, while the opposite polarization is repelled and forms itinerant states whose density is reduced near the skyrmion center. The estimate of $E_T$ in the following subsection neglects electron-electron interaction altogether. We now evaluate $E_T$ for the two skyrmion profiles discussed above.

\subsection{Belavin-Polyakov skyrmion}
\subsubsection{Harmonic states}

We start by computing the contribution of harmonic states in the limit of large skyrmion radius $R\gg \xi$. Treating the valence band maximum as energy reference, we rewrite Eq. (\ref{Energy_OH_BP}) as 
\begin{equation}
	\varepsilon_m^n = \varepsilon_0^0 - \left(\frac{1}{2} + n\right) \hbar\omega _y - \left(\frac{1}{2} + m\right) \hbar\omega_x ,
\end{equation} 
with $\varepsilon_0^0 = \Delta \frac{\xi}{R}$. Which allows us to write
\begin{equation}
	E_T = \sum_{m,n} \varepsilon_m^n = \sum_{n=0}^{N_y} \sum_{m=0}^{M_n} \left(\varepsilon_0^n - \left(\frac{1}{2} + m\right) \omega_x\right) , \\
\end{equation}
with $N_y = \frac{\varepsilon_0^0}{\hbar\omega_y}$ and $M_n = \frac{\varepsilon_0^n}{\hbar\omega_x}$. Since $N_y \gg 1$, this yields 
\begin{equation}
		E_T \simeq \frac{\left(\varepsilon_0^0\right)^3}{6 \hbar^2\omega_x\omega_y} = \frac{\Delta}{6}\frac{(\xi/R)^3}{2\sqrt{\epsilon}(\xi/R)^3} \frac{1}{\sqrt{1+\xi/2R}}
  , \label{E_tot limite harmonique}
\end{equation} 
which implies asymptotic behaviour 
\begin{equation}
	\lim_{R\to\infty} E_T(R) \simeq E_0 ,\;\; \text{with} \;\; E_0 \equiv \frac{\Delta}{12\sqrt{\epsilon}} . \label{E_T assymptotic BP}
\end{equation}
Notice that, at large $R$, the increasing number of bound states and their decreasing characteristic energy compensate each other. Also notice that $E_0 \gg \Delta$ emerges as a characteristic scale, defining the contribution of bound states to the skyrmion energy. Note that the estimate above treats bound states as harmonic (equidistant) and underestimates their number, thus giving a lower bound of the bound-state contribution $E_T$ to the skyrmion energy. 

We now evaluate $E_T$ for intermediate radia \mbox{$R_c \ll R \lesssim \xi$}.

\subsubsection{Contribution of deep bound states}

\begin{figure}[ht]
	\centering \includegraphics[width=\linewidth]{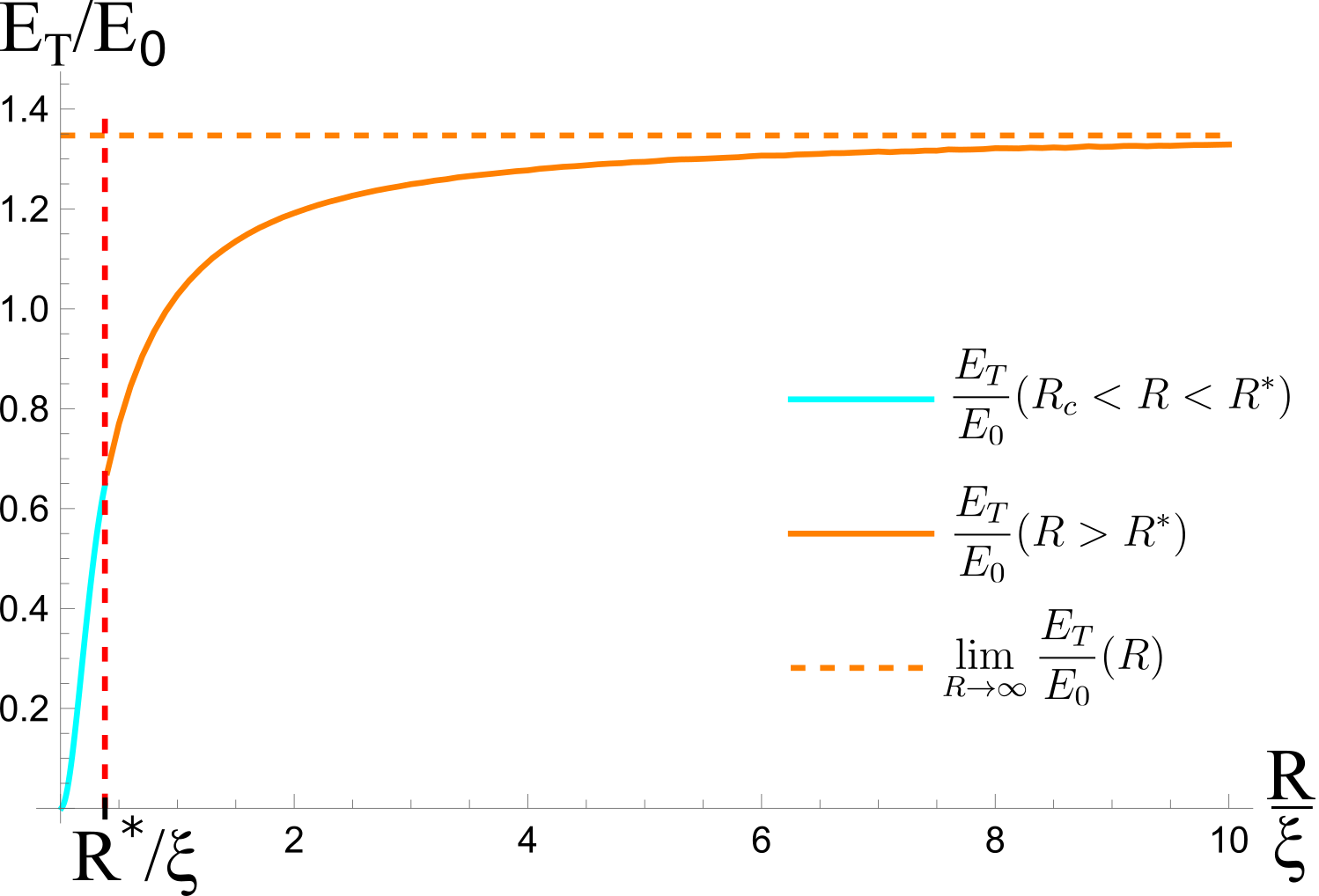} 
	\caption{Numerical estimate of the bound-state contribution to the skyrmion energy $E_T$, shown in  units of $E_0 = \frac{\Delta}{12\sqrt{\epsilon}}$. Orange line corresponds to Eq. (\ref{E R > R* }), where the sum is calculated numerically for the first seven excited states, the maximum of the 7$^{\text{th}}$ level energy being already smaller than $4.10^{-3}\Delta$ for this radius range. The apparent asymptotic value, close to $1.34 E_0$, is of the same order of magnitude but slightly larger than the one obtained with harmonic approximation (\ref{E_T assymptotic BP}), as expected. The blue part is calculated for radius smaller than the crossing radia $R< R^*$ using Eq.~(\ref{Energie_BP_R<R*}).}
	\label{fig:E_tot_BP_numerique}
\end{figure}

 In the following section, we focus on the density of in-gap states since, in the first approximation, the sum over energies does not depend on the fine structure of  energy levels. This amounts to ignoring the perturbation of harmonic structure, discussed in subsection \ref{subsec : Perturbation theory}. In this context, the physics picture is roughly the one presented in Fig.~\ref{fig: Structure interne bande}. Following the steps of Chap.~\ref{sec: BP Deep bound states} and assuming that, in the intermediate regime $R \gtrsim R^*$, the harmonic description holds only for the $x$ direction, this yields 
\begin{equation}
	E_T \simeq \frac{1}{2\hbar\omega_x}\sum_n (\varepsilon_n^0)^2 \label{E R > R* } 
\end{equation}
which can be evaluated numerically, as shown in Fig. \ref{fig:E_tot_BP_numerique}.

\begin{figure}[ht]
	\centering \includegraphics[width=0.9\linewidth]{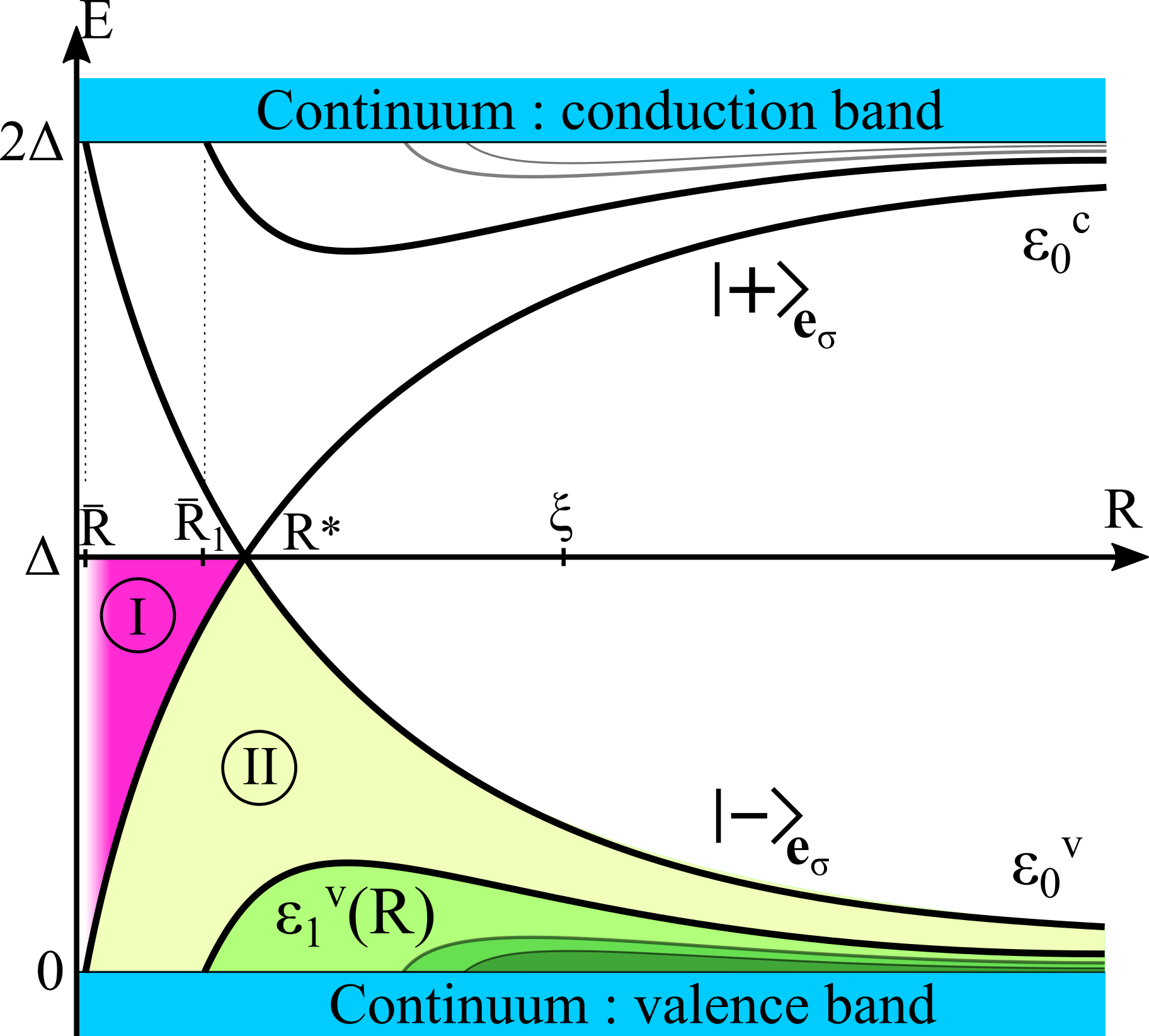} 
	\caption{Schematic representation of occupied states at half filling. Occupied ``bands'' are coloured. The top of the valence band is taken as zero energy, as for calculations made in this chapter. The bound state spectrum remaining approximately equidistant for $R \sim R^*$ allows us to approximate the density of states as being constant as a function of energy inside the gap, and equal to $1/\hbar\omega_x$ in region \textbf{II}. In the region \textbf{I}, where valence-band states $\varepsilon_v^0$ mix with conduction-band states $\varepsilon_c^0$, the density of states is expected to be about $2/\hbar\omega_x$. Note that the length $R_c$ is not shown here, since it is vanishingly small compared with the coherence length $\xi$, as is the loss radius $\bar{R}$. 
 }
	\label{fig:Occupied_states_BP_Rint}
\end{figure}

As the skyrmion radius decreases below $R^*$, the bound states produced by valence and conduction bands begin to hybridize, complicating the problem. However, for $R \gg R_c$, the `fine-structure' levels produced by quantization of motion along $x$ remain roughly harmonic (equidistant). That is, each state of the `gross' structure, labelled relative to the $y$ coordinate, can be seen as having a `fine'-structure band with a density of states
\begin{equation}
	\nu_0(R) = \frac{1}{\hbar \omega_x} = \frac{ E_0}{\sqrt{2}\Delta^2}\left(\frac{R}{\xi}\right)^{3/2}\left(1+\frac{\xi}{2R}\right)^{-1/2} \label{DOS}
\end{equation}
attached to it. In this picture, for $R \geq \bar{R}_1$ (with $\bar{R}_1$ being the radius where the first excited bound state vanishes) the band $\varepsilon_1^v$ associated with first excited state can be considered completely filled. The bands $\varepsilon_0^v$ and $\varepsilon_0^c$ are filled up to energy $\Delta$, see Fig. \ref{fig:Occupied_states_BP_Rint}. The resulting total energy cost can thus be estimated as 
\begin{equation}
	\begin{split}
		E_T &\sim \frac{\nu}{2} \left(\varepsilon_1^0\right)^2 + \nu \int_0^\Delta \varepsilon d\varepsilon + \nu\int _{2\Delta-\varepsilon_0^0}^\Delta \varepsilon d\varepsilon \\
		&\sim \frac{\nu}{2}\left(\varepsilon_1^0\right)^2  + \nu \Delta^2 \left( -1 + 2\frac{\varepsilon_0^0}{\Delta} -\frac{1}{2} \left(\frac{\varepsilon_0^0}{\Delta} \right)^2\right). \label{Energie_BP_R<R*}
	\end{split}
\end{equation}
For $R < R_1$ this result holds by simply taking $\varepsilon_1^0 = 0$, the corresponding plot is given in Fig. \ref{fig:E_tot_BP_numerique}.   

For $R < R_c$, the numerical results in Fig. \ref{fig:E_tot_BP_numerique} can no longer be trusted. However, at $R < \bar{R}$, there are no bound states any more, and thus the bound-state contribution to the skyrmion energy goes to zero at $R \to \bar{R}$. 
Therefore, to minimize the bound-state contribution to its energy, a BP skyrmion shall shrink to a radius comparable to $\bar{R}$.
Note that these results are only qualitative, but the overall picture of the energy cost $E_T(R)$ should be captured by the above derivation. We now turn to estimating the bound-state contribution to the energy of a DW skyrmion.

\subsection{Domain-wall skyrmion}
\begin{figure}[ht]
	\centering \includegraphics[width=\linewidth]{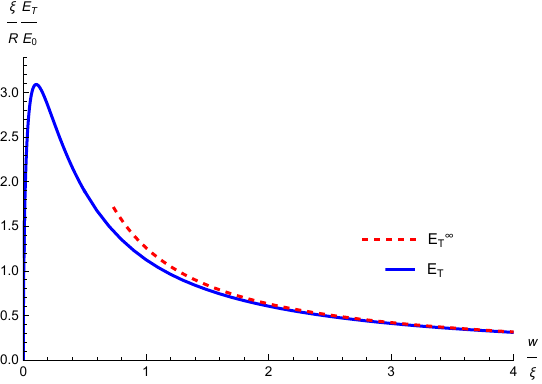} 
	\caption{The bound-state contribution to the DW skyrmion energy $E_T$, shown in units of $E_0 \frac{R}{\xi}$ as a function of the wall width $w$. The asymptotic behaviour is $E_T \simeq 1.26 E_0 \frac{R}{w}$, with a coefficient slightly greater that the factor $1$ predicted within harmonic approximation, as expected.}
	\label{fig: E_tot(w,R) DWB} 
\end{figure}
The shallow bound-state limit $w \gg \xi$ can be treated as for the BP profile, with the help of Eqs. (\ref{Enm DW}) and (\ref{E_tot limite harmonique}), yielding 
\begin{equation}
	E_T \underset{w \gg \xi}{\sim} 2\frac{\varepsilon_0^3}{6\hbar^2 \omega_x\omega_y}
 = E_0\frac{R}{w} ,
 \label{eq:ET_for_shallow_DW}
\end{equation}
where the factor of 2 above accounts for the two potential wells generated by the skyrmion. 
 The bound-state contribution $E_T$ to the skyrmion energy thus decreases with $w$ increasing relative to the DW skyrmion radius $R$, but only for $w \ll R$. Meaning that (for a given $R$) $E_T$ has a lower bound of the order of $E_0$. For the DW profile,  there is no crossing of bound states that emerge from the conduction- and valence bands. Therefore, the energy cost for $w \lesssim \xi$ can be estimated as we did it for the BP profile for $R > R^*$ (see Eq.~(\ref{E R > R* })), leading here to
 \begin{equation}
 	E_T \simeq \frac{2}{2\hbar\omega_x}\sum_n (\varepsilon_n^0)^2 = 12\sqrt{2}E_0\frac{R}{\xi}\sqrt{\frac{w}{\xi}} \sum_n \frac{(\varepsilon_n^0)^2}{\Delta^2}. \label{E_T_DW_petit w}
 \end{equation} 
 Estimating the above sum numerically yields Fig. \ref{fig: E_tot(w,R) DWB}. The energy cost $E_T$ tends to zero as $w \rightarrow 0$ and, naturally, grows linearly with the DW skyrmion radius $R$. 
 
 In a way, the above behaviour of $E_T(R,w)$ favors the skyrmion shrinking to reduce both its radius $R$ and the wall width $w$. In other words, the bound-state contribution favours compact profiles similar to the BP one, producing a single potential well.

\subsection{Straight domain wall}
\label{subsec:StDW}

 \begin{figure}[ht]
	\centering \includegraphics[width = \linewidth]{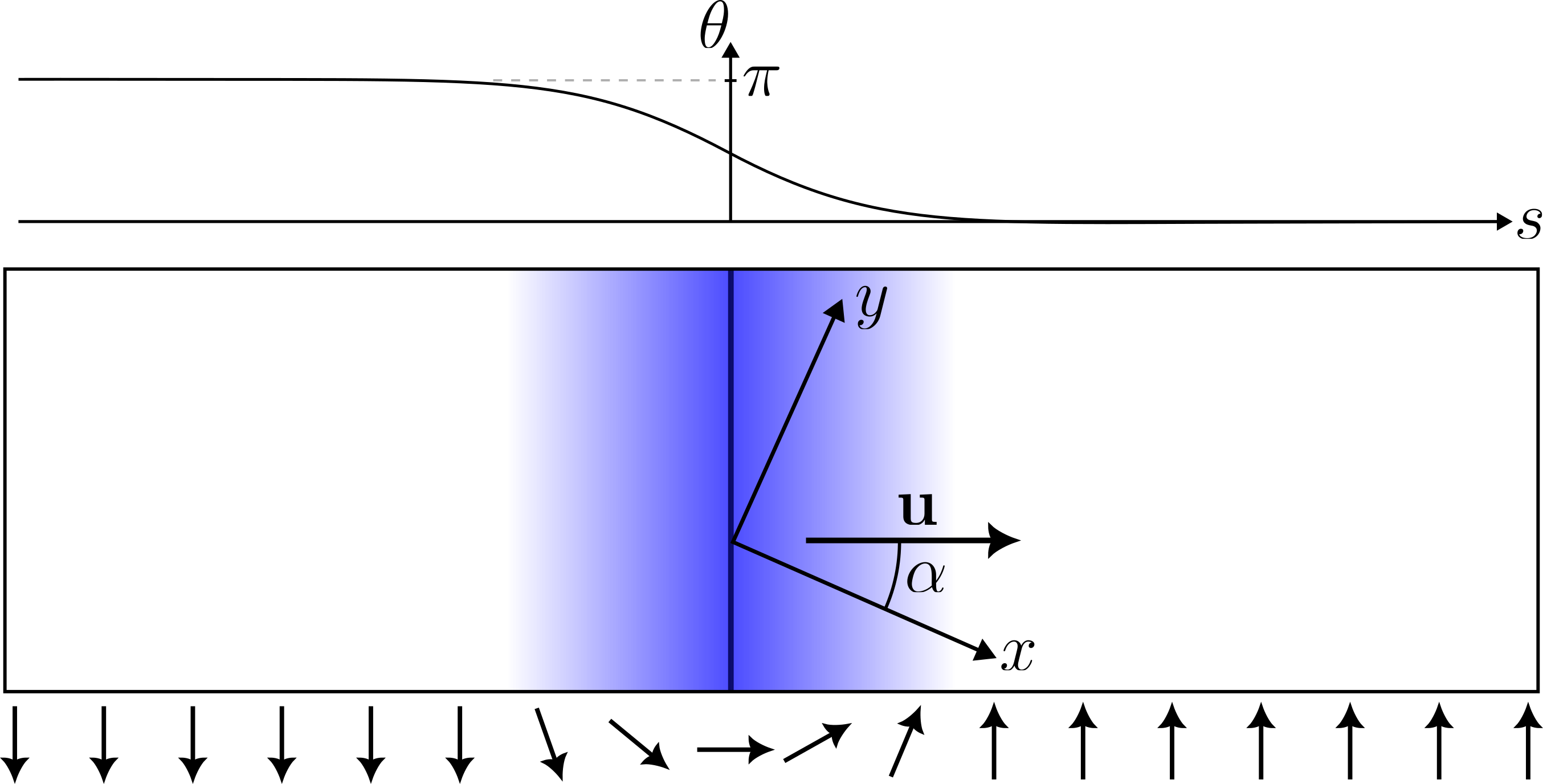} 
	\caption{Straight domain wall at an angle $\alpha$ to the 
 $x$-axis in real space, the axis defined for a given valley 
 as shown in Fig.~\ref{fig: MBZ}. Unit normal $\mathbf{u}$ 
 to the wall points at the same angle $\alpha$ with respect 
 to the $x$ axis. Polar angle $\theta$ varies along $\mathbf{u}$ 
 as a function of $s$ in Eq. (\ref{def SDW}).  
Angle $\alpha$ governs the bound-state formation: 
 the attractive potential $v (\mathbf{A}_y\cdot \bm{\sigma})$ of Eq. (\ref{eq:A4straightDW})
 is the strongest for $\alpha = \pi/2$, that is for $\theta$ dependent only on $y$ while the wall stretches along $x$. By contrast, at $\alpha = 0$ attraction ceases to exist: $v (\mathbf{A}_y\cdot \bm{\sigma} ) \equiv 0$ for $\theta$ dependent only on $x$ while the wall stretches along $y$.
 }
	\label{fig: SDW} 
\end{figure}
Now we consider an electron in the presence of a straight domain wall. The problem is interesting for several reasons: firstly, in its own right. Secondly -- because it is quasi-exactly solvable. And, last but not least -- since it allows one to gain insight into the origin of the peculiar form of the effective potential in Eq. (\ref{vA para DW}) (see Fig. \ref{fig: Potentiel_DW}) as well as into the effect of electron bound states on the transition to the modulated phase \cite{Bogdanov_1989}.

Orientation of a straight domain wall in real space is defined by an in-plane unit vector 
$\mathbf{u} = (\cos\alpha, \sin\alpha,0)^t$, orthogonal to the wall, see Fig.~\ref{fig: SDW}, while its profile is described by 
\begin{equation}
	\theta(s) =  2\arctan \left(e^{-s} \right), \hspace{0.7cm} s \equiv \frac{\cos(\alpha) \, x + \sin(\alpha)\, y}{w}. \label{def SDW}
\end{equation}
According to (\ref{definition A_i}), it produces a simple gauge field 
\begin{equation}
	\mathbf{A}_i\cdot \bm{\sigma} = \frac {1}{2} 
 (\mathbf{e}_\sigma \cdot \bm{\sigma}) \partial_i\theta \, 
\end{equation}
where $\mathbf{e}_\sigma = ( -\sin \phi, \cos\phi, 0)^t$ with $\phi = \alpha + \gamma$. Parameter $\gamma$ is similar to the helicity of a skyrmion, and defines bound-state polarization. The only potential term reads
\begin{equation}
	v ( \mathbf{A}_y\cdot \bm{\sigma} ) = 
 ( \mathbf{e}_\sigma \cdot \bm{\sigma}) 
 \frac {\Delta}{2} \frac{\xi}{w} \sin\alpha \,\text{sech}\,s\, ,
 \label{eq:A4straightDW}
\end{equation}
it vanishes for $\sin \alpha = 0$. This is hardly surprising: the presence of $\mathbf{A}_y = - i \hbar U^\dagger (\partial_y U)$ in the l.h.s. above means that, to produce attraction and bound states for electrons in the given valley, the N\'eel order parameter $\mathbf{n}$ must depend on $y$. Viewing a large-radius DW skyrmion as a circle, and approximating it as a regular inscribed polygon with straight domain-wall segments, we see from Eq. (\ref{eq:A4straightDW}) that the attraction is the strongest on segments that are nearly normal to the $y$ axis -- indeed, as we already saw in Fig. \ref{fig: Potentiel_DW}.

The limit $w\gg\xi$ can again be treated with effective Hamiltonian (\ref{H_0}) in the harmonic approximation, 
yielding the low-lying spectrum 
\begin{equation}
	\begin{split}
		\varepsilon_k^n =&\; - \Delta \frac{\xi}{w} \sin\alpha + \left(\frac{1}{2} + n\right)\frac{\hbar}{w} \sqrt{\frac{\Delta\xi\sin\alpha}{m_s w}}\\ &+ \left(\frac{1}{2m_q} - \frac{m_s}{8M^2}\right)\frac{\hbar^2k^2}{w^2}.
	\end{split}
\end{equation}
We introduced here effective masses 
\begin{equation}
	\begin{split}
		\frac{1}{m_s} = \frac{\cos^2\alpha }{m_x} + \frac{\sin^2\alpha}{m_y^*}, \hspace{0.5cm} \frac{1}{m_q} = \frac{\sin^2\alpha }{m_x} + \frac{\cos^2\alpha}{m_y^*},\\ \frac{1}{M} = \sin2\alpha \left(\frac{1}{m_y^*}-\frac{1}{m_x}\right) \hspace{1.5cm}
	\end{split}
\end{equation}
along pertinent directions, and the wave vector $k$ along the domain wall. 
The energy cost of a domain wall in this limit can now be evaluated as 
\begin{equation}
	E_T = -\sum_n \int _{\varepsilon_0^n}^0 \nu(\varepsilon) \varepsilon d\varepsilon \simeq E_{DW} \sum_n \left( \frac{|\varepsilon_0^n|}{\Delta} \right)^{3/2} 
 \label{eq: ET SDW}
\end{equation}
with $\nu$ the DOS associated with delocalized states along the wall, and $E_{DW} = \frac{16 \sqrt{2}}{\pi} E_0 \frac{L}{\xi} \sin\alpha$ the characteristic energy of a DW. A simple calculation shows that for $w \gg \xi$ this expression tends to
\begin{equation}
		E_T \simeq \frac{32\sqrt{2}}{5\pi}\sin^2\alpha E_0\frac{L}{w}, 
  \label{eq:ET-straight-DW}
\end{equation}
showing that the energy cost grows linearly with the domain wall length $L$, indeed as expected. 

\begin{figure}[ht]
    \centering
    \includegraphics[width = \linewidth]{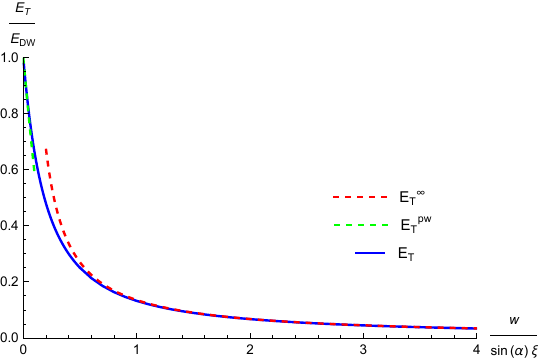}
    \caption{Numerical estimation of the energy $E_T$ of a straight DW, performing the sum (\ref{eq: ET SDW}). The result is given in units of $E_{DW}$, the natural energy unit for a straight DW. The asymptotic behaviour for $w \geq \xi$ is in $\left(\frac{w}{\sin (\alpha) \xi}\right)^{-1}$, see dashed red curve, as expected from the harmonic approximation (\ref{eq:ET-straight-DW}). The limit of $E_T$ for $w \ll \xi$ is finite and precisely equal to $E_{D}$, see main text. The dashed green curve give the estimation of the sum (\ref{eq: ET SDW}), by fitting linearly $\varepsilon_0^0(w)$.  }
    \label{fig: Etot SDW}
\end{figure}

For $w \lesssim \xi$, Eq. (\ref{Eq générale phi}) can be applied directly: 
note that this time there is no need to neglect the term $vA_y^z \sigma^z$ in $\hat{\bm{\gamma}}_-$ since now it is identically equal to zero. Eq. (\ref{Eq générale phi}) is equivalent to the one for a DW skyrmion, up to a redefinition of the coherence length $\xi \to \sin\alpha \, \xi$. The r.h.s. of Eq. (\ref{eq: ET SDW}) can then be evaluated using the energy derived numerically for a circular skyrmion, the result is given in units of $E_{DW}$ in Fig.~\ref{fig: Etot SDW}. The limit $w \geq \xi$ shows a $\left(\frac{w}{\sin (\alpha) \xi}\right)^{-1}$ behaviour as expected. Note that the total energy cost of a DW always varies linearly with its length $L$, as does $E_{DW}$. In the region $w \ll \xi$, there is a single non-zero term in the sum (\ref{eq: ET SDW}) that tends toward 1, corresponding to a unique occupied bound-state band. The bound-state contribution to the energy of the domain wall is then expected to converge toward $E_{DW}$ as $w$ goes to zero.\\

Not surprisingly, for the wall orientation $\alpha = 0$, the $E_T$ in Eq. (\ref{eq:ET-straight-DW}) vanishes, simply because the N\'eel order parameter does not vary along $y$, hence no attractive potential and no bound states.
However, this is true only for the valley $\Sigma_i$ that we considered to define $\alpha$: for the next valley, $\Sigma_{i+1}$, one has $\alpha_{i+1} = \alpha_i + \frac{\pi}{2}$. That is, taking into account contributions from all the four $\Sigma$-valleys, and considering the $w \gg \xi$ limit, $E_T$ would contain two terms with angle dependence $\sin^2 \alpha$ and two terms with $\sin^2\left(\alpha+\frac{\pi}{2}\right) = \cos^2\alpha$, which yields 
\begin{equation}
    E_T \simeq \frac{64\sqrt{2}}{5\pi} E_0 \frac{L}{w} ,
    \label{eq:bs-cost-straight}
\end{equation}
meaning that the bound-state contribution to the domain-wall energy from all the four $\Sigma$ valleys does not depend on the orientation of the wall. \\

Equation (\ref{eq:bs-cost-straight}) and its counterpart for a circular DW skyrmion, 
Eq. (\ref{eq:ET_for_shallow_DW}), imply that skyrmion-electron bound states strongly influence the 
transition to the modulated phase \cite{Bogdanov_1989}. 
To see this, recall that the bound-state contribution to the skyrmion energy complements the purely magnetic part $E[\mathbf{n}]$ 
of the energy of a texture \cite{Bogdanov_1989}:
\begin{equation}
        E[\mathbf{n}] = \int d^2 r\left[ J\left(\bm{\nabla} \mathbf{n}\right)^2 + \frac{K}{a_0^2} (1-n_z^2) + \frac{D}{a_0} \epsilon_{DM}  \right] ,
    \label{eq: DW mangetic energy} 
\end{equation}
where $J$ is stiffness, $K$ the easy-axis anisotropy, $D$ the Dzyaloshinskii–Moriya coupling, and the Dzyaloshinskii–Moriya energy density  $\epsilon_{DM}$ depends linearly on $\nabla \mathbf{n}$. 
For a domain wall of length $L$ and width $w$, one has $| \bm{\nabla} \mathbf{n} | \sim 1/w$ within the range of about $w$ from the center of the wall and zero otherwise, while $(1-n_z^2) \sim 1$ within the same range of about $w$, and zero elsewhere. With this estimate in mind, na\"ive integration of the different terms in Eq. (\ref{eq: DW mangetic energy}) along the wall and transversely to it yields the following characteristic dependence of $E[\mathbf{n}]$ on $L$ and $w$:
\begin{equation}
        E[\mathbf{n}] 
        \sim L \left[ \frac{J}{w} + \frac{K}{a_0^2} w - \frac{D}{a_0}\right] .
        \label{eq:DW-scaling}
\end{equation}
Eq. (\ref{eq:DW-scaling}) is written up to numerical coefficients of the order of unity, dependent on the details of the domain wall profile, just as the coefficients in Eqs. (\ref{eq:bs-cost-straight}) and (\ref{eq:ET_for_shallow_DW}). These coefficients do not affect the conclusion below. 
The first two terms in Eq. (\ref{eq:DW-scaling}) define the optimal domain wall width $w_0 = a_0 \sqrt{\frac{J}{K}}$ 
that minimizes $E[\mathbf{n}]$, thus producing the characteristic dependence of the domain wall energy on $J$, $K$ and $D$:
\begin{equation}
E (w_0) \sim \frac{L}{a_0} \left[ \sqrt{JK} - D \right] .
\end{equation}
A more accurate calculation yields the energy $E (w_0)$ becoming negative at 
\begin{equation}
\label{eq:Bogdanov}
D_c = \frac{4}{\pi} \sqrt{JK} , 
\end{equation}
marking the transition to a modulated state \cite{Bogdanov_1989}. 

\begin{figure}[ht]
    \centering
    \includegraphics[width = 0.8\linewidth]{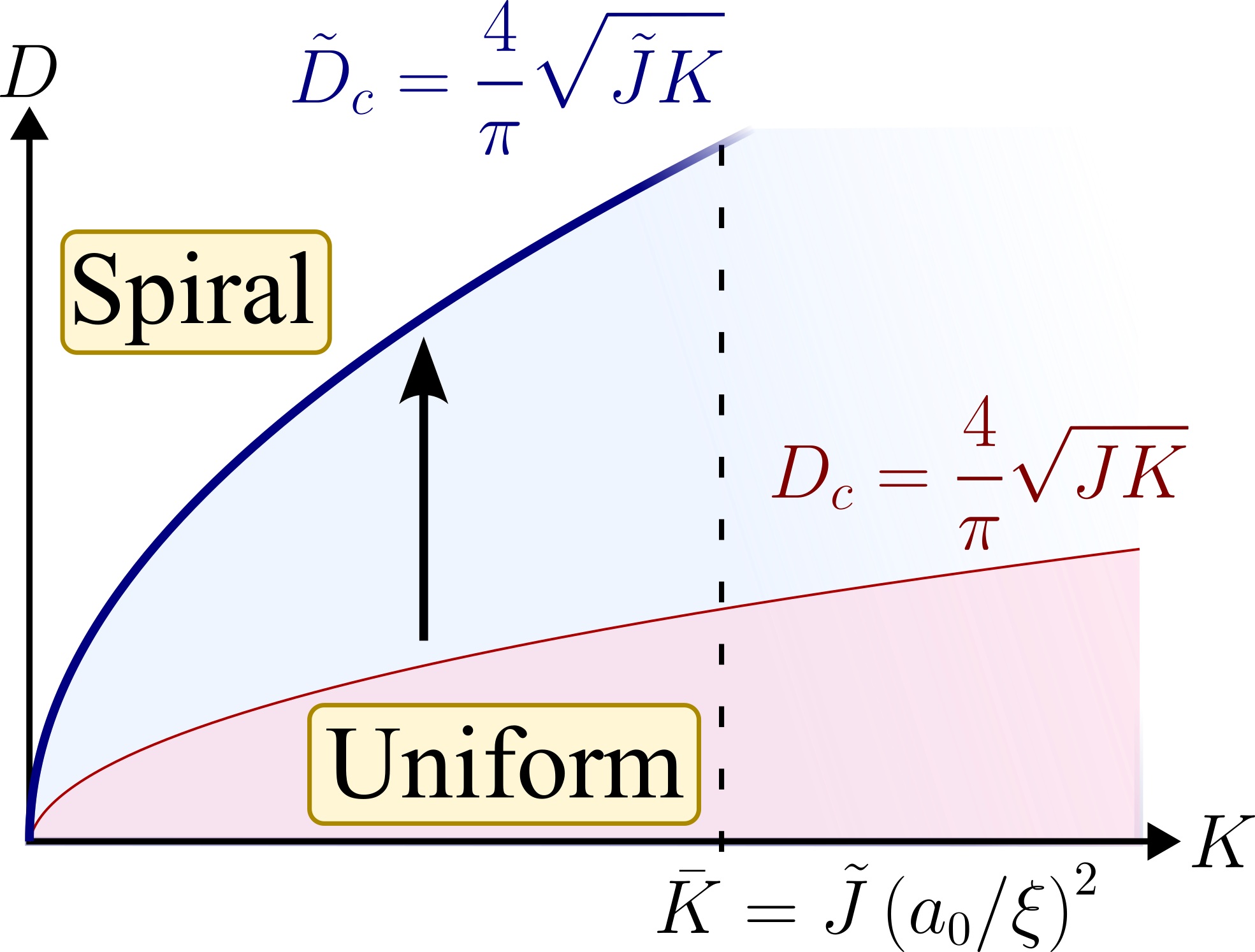}
    \caption{The effect of skyrmion-electron bound states on the phase diagram: the critical value $D_c [J , K]$, marking the transition between the uniform and the modulated phases, increases from $D_c=\frac{4}{\pi} \sqrt{JK}$ 
    to $\tilde{D}_c=\frac{4}{\pi} \sqrt{\tilde{J}K}$ as per Eq. (\ref{eq:Jrenorm}), while keeping its $\sqrt{K}$ dependence. Notice that this picture holds for $w \gtrsim \xi$, which limits it to small values of $K \lesssim \tilde{J} \left[ \frac{a_0}{\xi}\right]^2$. 
    }
    \label{fig: Jrenorm}
\end{figure}

Now, comparison of Eq.  (\ref{eq:DW-scaling}) with Eqs. (\ref{eq:bs-cost-straight}) 
and  (\ref{eq:ET_for_shallow_DW}) shows that the bound-state contribution $E_T$ to the domain-wall energy 
renormalizes and, in fact, makes a dominant contribution to stiffness $J$ in Eqs. (\ref{eq:DW-scaling}) 
and (\ref{eq:Bogdanov}) as per 
\begin{equation}
J \rightarrow \tilde{J} = J + \frac{\Delta}{\sqrt{\epsilon}} 
 = J + \sqrt{\frac{m_x}{m_y}} \sqrt{\varepsilon_F \Delta} .
 \label{eq:Jrenorm}
\end{equation}
The significant upward renormalization of $J$ in Eq. (\ref{eq:Jrenorm}) increases the domain wall width. Even more importantly, it  substantially increases $D_c [J , K]$ in Eq. (\ref{eq:Bogdanov}) while keeping its $\sqrt{K}$ scaling, and thus extends the stability range of the uniform state, as shown in Fig. \ref{fig: Jrenorm}. Note that this picture holds for $w \gtrsim \xi$, that is for sufficiently small values of $K \lesssim \tilde{J} \left[ \frac{a_0}{\xi}\right]^2$.

Numerically, how large a renormalization may arise from Eq. (\ref{eq:Jrenorm})? 
As a concrete example, consider optimally-doped high-temperature superconductor NCCO (Nd$_{1.85}$Ce$_{0.15}$CuO$_4$). 
The presence of N\'eel order in this compound remains controversial \cite{Luke_1990,Mang_2004,Motoyama_2007,Saadaoui_2015}, 
yet a number of experiments \cite{Yamada_2003,Kang_2005,Yu_2007,Dorantes_2018,RR_Zeeman_SOC_2021} 
report static or quasi-static N\'eel order. The electron subsystem of NCCO is nearly two-dimensional, 
with the low-temperature ratio of the $c$-axis and $ab$-plane resistivities of about $10^3$ to $10^4$ \cite{Armitage_2010}.
The two-dimensional band structure of NCCO involves small carrier pockets at points $\Sigma$, 
with anisotropic dispersion as in Fig. \ref{fig: MBZ}. Assuming quadratic dispersion 
$\frac{p_x^2}{2 m_x} + \frac{p_y^2}{2 m_y}$, the mass anisotropy can be read off ARPES maps 
in Refs \cite{Armitage_2002,He_2019} as $m_x / m_y \sim 25$. The $\varepsilon_F / \Delta \sim \xi / a$ 
ratio can be evaluated, knowing the NCCO lattice constant  $a \approx 3.95${\AA} \cite{Armitage_2010}, 
and estimating $\xi \sim 9$ nm -- for instance, from magnetic quantum oscillation data \cite{RR_Zeeman_SOC_2021}. 
As a result, for $\tilde{D}_c / D_c$ in Fig. \ref{fig: Jrenorm}, with numerical parameters of NCCO we find  
$$
\frac{\tilde{D}_c}{D_c} \sim \sqrt{\frac{\tilde{J}}{J}} \sim \sqrt[4]{\frac{m_x}{m_y} \frac{\varepsilon_F}{\Delta}} \sim 5 .
$$
The bound-state contribution to the energy of a domain-wall texture in Eqs. (\ref{eq:bs-cost-straight}) and (\ref{eq:ET_for_shallow_DW})  proves large compared with the `magnetic' contribution of Eq. (\ref{eq: DW mangetic energy}). This is in contrast with what one finds for electron states, bound to a vortex in a type-II superconductor \cite{DeGennes_Book, tinkham2004}: in the latter problem, the gradient contribution to the vortex energy has the form $\left( \frac{\phi_0}{4 \pi \lambda} \right)^2 \ln \frac{\lambda}{\xi}$, where $\xi$ is the superconducting coherence length, and $\lambda \gg \xi$ the penetration depth. The electron states, bound to the vortex core, only correct the large argument of the logarithm by a coefficient of the order of unity \cite{DeGennes_Book, tinkham2004}, thus making a parametrically small contribution to the energy of the vortex. 

\section{Discussion}
\label{sec: Discussion}

\subsection{Validity of static-skyrmion approximation} 
\label{subsec: Skyrmion eigenmodes}

Above, all the textures were implicitly treated as static input 
to the electron problem. This assumption must be taken with caution, 
since an isolated antiferromagnetic skyrmion possesses its own 
excitation modes \cite{Kravchuk_2019} that couple to the bound electron. 
To justify viewing the skyrmion as static, these eigenmodes 
must have frequencies far below those of electron bound states. 
The proper skyrmion frequencies are closely bound from above 
\cite{Kravchuk_2019} by the gap $\hbar \Omega_0$ in the 
magnon spectrum, which is about $\hbar \Omega_0 \sim \sqrt{2JK}$ 
\cite{Rezende2019, Kittel_1951} for systems where 
single-ion anisotropy $K$ is much smaller than stiffness $J$. 
Thus, it is $\Omega_0$ that must be compared with 
characteristic electron frequencies, whose scales are different for 
the `gross' and `fine' structures of bound-electron levels. 

For the BP profile, the `gross' structure is defined by 
the frequency $\hbar \omega_y \sim \Delta \left( \frac{\xi}{R} \right)^{3/2}$
in the first term of Eq. (\ref{Energy_OH_BP}). Inequality 
$\omega_y \gg \Omega_0$ limits the validity of the 
static-texture approximation to not-too-large skyrmions 
\begin{equation} 
\nonumber
  R \ll  R_y^{BP} = \xi \left( \frac{\Delta^2}{JK}\right)^{1/3}  .
\end{equation}
At the same time, consistency with the effective-Hamiltonian approach  
requires that the bound states remain shallow. According to 
Eq. (\ref{Energy_OH_BP}), this implies $R \gg \xi$. 
Therefore, the upper bound $R_y^{BP}$ above must be large 
compared with $\xi$, which thus demands \cite{Davier_2023}
\begin{equation}
\label{ineq:K4BP-gross}
  \frac{K}{J} \ll \left( \frac{\Delta}{J}\right)^2 .
\end{equation}
The r.h.s. of the inequality (\ref{ineq:K4BP-gross}) may be of the order 
of unity or smaller: recall that, for a lattice model, $\Delta$ scales 
linearly with the lattice spin $S$ (ordered moment), whereas $J \sim S^2$. Nevertheless, for a not-too-large $S$, inequality (\ref{ineq:K4BP-gross}) 
simply requires that $K/J$ be small against unity.

For the `fine' structure of the bound-state spectrum, the condition 
to meet for the BP profile is more stringent: now it is the `low' frequency 
$\omega_x$ in the second term of Eq. (\ref{Energy_OH_BP}) that must be 
large relative to $\Omega_0$: 
\begin{equation}
    \hbar \omega_x \sim \Delta \sqrt{\epsilon} \left(\frac{\xi}{R}\right)^{3/2} \gg \sqrt{2JK}. 
\end{equation}
This limits the validity of the static-texture approach to
\begin{equation}
\nonumber
  R \ll  R_x^{BP} = \epsilon^{1/3} \xi 
  \left(\frac{\Delta^2}{JK}\right)^{1/3}  .
\end{equation}
Again, for consistency with the effective-mass approximation, 
$R_x^{BP}$ must be large against $\xi$, which thus demands  
\begin{equation}
\label{ineq:K4BP-fine}
\frac{K}{J}  \ll 
\epsilon \left( \frac{\Delta}{J} \right)^2 .
\end{equation}
Inequality (\ref{ineq:K4BP-fine}) is similar to (\ref{ineq:K4BP-gross}), 
but has an extra factor of $\epsilon = \frac{m_y}{m_x} \frac{a}{\xi} \ll 1$ in the right-hand side.

Numerically, how stringent are inequalities (\ref{ineq:K4BP-gross}) and (\ref{ineq:K4BP-fine})? 
As in the immediately preceding section, we turn to NCCO as a concrete example. The values used 
in Subsection \ref{subsec:StDW} yield $\epsilon \sim 2 \cdot 10^{-3}$. As a result, the `gross' 
structure of electron states, bound to a not-too-large BP skyrmion, can be treated in the 
static-skyrmion approximation for the $K/J$ ratio sufficiently small compared with unity. 
By contrast, treating the fine structure of levels in the same approximation requires 
a much smaller anisotropy $K/J \ll 10^{-3}$. 

For a domain-wall (DW) profile, the wall width $w$ is defined by anistropy 
as per $w \sim a_0 \sqrt{\tilde{J}/K}$, see Sec. \ref{subsec:StDW}. Just as 
for the BP skyrmion, one constraint on the validity of our results stems 
from Eq. (\ref{Enm DW}) and from the requirement that the bound states 
remain shallow, which implies $w \gg \xi$. 
The latter inequality can be recast as 
\begin{equation*}
    \frac{K}{J} \ll \frac{\tilde{J}}{J} \left(\frac{a_0}{\xi}\right)^2 . 
\end{equation*}
Using Eq. (\ref{eq:Jrenorm}) and treating the second term in its r.h.s. 
as dominant, one finds 
\begin{equation} 
\label{ineq:DW-shallow}
\frac{K}{J} \ll \frac{\Delta}{J} \sqrt{\frac{m_x}{m_y}} 
\left(\frac{a_0}{\xi}\right)^{3/2} .
\end{equation}
At the same time, the validity of the static-skyrmion approximation 
is limited by the condition that the characteristic electron frequency 
be large compared with that of skyrmion eigenmodes. 

For the `gross' structure of electron levels this means that the 
scale $\Delta (\xi / w)^{3/2}$ of the `high' oscillator frequency 
in the first term of Eq. (\ref{Enm DW}) must be large compared with 
the gap $\hbar \Omega_0 \sim \sqrt{2JK}$ in the magnon spectrum. 
Taking into account the characteristic domain wall width 
$w \sim a_0 \sqrt{\tilde{J}/K}$, this limits the single-ion 
anisotropy $K$ from \textit{below} rather than from above:
\begin{equation}
\label{ineq:DW-gross-static}
\frac{K}{J} \gg \frac{J}{\Delta} \left( \frac{a_0}{\xi} \right)^{9/2} 
\left( \frac{m_x}{m_y} \right)^{3/2} .
\end{equation} 
Together, the inequalities (\ref{ineq:DW-shallow}) 
and (\ref{ineq:DW-gross-static}) read 
\begin{equation}
\label{ineq:DW-gross-combined}
\frac{J}{\Delta} \left( \frac{a_0}{\xi} \right)^{9/2}
\left( \frac{m_x}{m_y} \right)^{3/2} 
\ll \frac{K}{J} \ll 
\frac{\Delta}{J} \sqrt{\frac{m_x}{m_y}} 
\left(\frac{a_0}{\xi}\right)^{3/2} .
\end{equation}
What numerical bounds can this produce? 
Again, for a rough estimate, consider NCCO as a 
concrete (even if not necessarily universal) example. 
The material parameters above yield 
\begin{equation*}
    10^{-4} \ll \frac{K}{J} \ll 5 \cdot 10^{-2}.
\end{equation*}
Notice that the upper-to-lower bound ratio in 
inequality (\ref{ineq:DW-gross-combined}) is 
\begin{equation*}
    \left( \frac{\Delta}{J} \right)^2 
\frac{m_y}{m_x} 
\left(\frac{\xi}{a_0}\right)^3 \gg 1 .
\end{equation*}
It contains (i) the $\left(\frac{\Delta}{J}\right)^2$ factor that is likely 
of the order of unity, (ii) the $\frac{m_y}{m_x}$ ratio that 
tends to be small albeit not parametrically so (about 0.04 for NCCO) 
and, finally, (iii) the factor $\left(\frac{\xi}{a_0}\right)^3$ 
that is indeed parametrically large (over $10^4$ for NCCO) 
and overwhelms all the rest.

To treat the `fine' structure of electron levels in the 
static-skyrmion approximation, one has to demand that the 
`low' frequency in the second term of Eq. (\ref{Enm DW}) 
be large compared with the gap $\hbar \Omega_0 \sim \sqrt{2JK}$ 
in the magnon spectrum:
\begin{equation} 
\Delta \sqrt{\epsilon} \frac{\xi^{3/2}}{w^{1/2} R} \gg \sqrt{JK} .
\end{equation}
With $w \sim a_0 \sqrt{\tilde{J}/K}$, this inequality 
translates into another \textit{upper} bound, but 
far more stringent than (\ref{ineq:DW-shallow}):
\begin{equation} 
\frac{K}{J} \ll 
\left( \frac{m_y}{m_x} \right)^{5/2} 
\left( \frac{a_0}{\xi} \right)^{1/2}
\left( \frac{\xi}{R} \right)^4 
\left( \frac{\Delta}{J} \right)^3 .
\end{equation}
Again assuming $\Delta/J \sim 1$, for NCCO the product of the 
first two factors above falls below $10^{-4}$. Now, recall that 
$\xi \ll w \ll R$. Even for $R/\xi \sim 10$, the factor 
$\left(\xi / R \right)^4$ above would then yield 
$K/J \ll 10^{-8}$. That is, realistically, for a DW skyrmion, 
fine structure of bound electron states can be treated 
in the static-skyrmion approximation only for perfectly isotropic 
magnets.

While the constraints of this section limit the validity of the static-skyrmion 
approximation as applied to shallow skyrmion-electron bound states, note that the 
\textit{appearance} of texture-bound electron states is a much broader phenomenon, 
extending to regimes where electron degrees of freedom and those of a texture must 
be treated on an equal footing. 


\subsection{Experimental signatures}
Combined with the appearance of skyrmion-electron bound states, doping a half-filled antiferromagnetic insulator turns a skyrmion into a charged particle. The latter can be detected and manipulated with the help of electric field. 

Even exactly at half-filling, resonance transitions between different bound-state levels would have a distinct frequency spectrum. The latter may involve transitions at $\hbar \omega < 2 \Delta$ (see Figs. \ref{fig: Structure interne bande} and \ref{fig: In-gap structure DW}), which could serve as a fingerprint of skyrmion-electron bound states. Away from half-filling, resonance transitions between the bound-state levels at frequencies $\hbar \omega \ll \Delta$ could serve as another spectroscopic hallmark. 
A study of such signatures will be the subject of forthcoming work. 

\subsection{Conclusion}
We examined a smooth texture in a N\'eel antiferromagnet with a specific location of the electron band extrema, and showed that it produces a peculiar spin-orbit coupling that has no analogue in a ferromagnet. The coupling has the scale $\hbar v / L$, where $v$ is the Fermi velocity of the underlying paramagnetic state, and $L$ is the relevant characteristic length scale of the texture. For topological textures such as skyrmions and domain walls, 
this coupling generates bound electron states in the gap of the electron spectrum.

Motivated by fundamental and technological interest alike, we focused on an interesting limit, 
where the proper frequencies of the skyrmion are small against those of the bound electron. 
In this limit, bound-state energies are sensitive to the skyrmion `shape'; 
the latter serves as a static input to the electron problem. 
We studied evolution of bound states with variation of the skyrmion 
profile for the Belavin-Polyakov and domain-wall skyrmions. 
Bound states prove to be energetically costly, and this 
(a) favors smaller BP skyrmions that produce no bound states, 
and (b) shifts the transition line between the 
uniform and the modulated states to substantially larger values 
of the Dzyaloshinskiy-Moriya coupling. 

With a major research effort focusing on magnetic topological textures, it is vital 
to understand their interplay with other degrees of freedom of the crystal, most notably 
with band electrons. We hope that our results contribute to a better understanding 
of magnetic textures and to putting them to use. 
\\

\section{Acknowledgement}
It is our pleasure to thank G. E. Volovik and V. Geshkenbein for enlightening discussions.

\appendix
\section{Comparison with the problem of a texture in a ferromagnet}
\label{Appendix : comparison with FM}

It is instructive to compare the problem we studied above with its ferromagnetic counterpart: 
band electron in the presence of a texture in a \textit{ferro}magnet. 
Smoothly varying as a function of coordinate $\bf{r}$, local magnetization $\bf{M_r}$ couples to the band electron spin $\bm{\sigma}$ 
via exchange term $J (\bf{M_r} \cdot \bm{\sigma})$ (cf. Subsection 
\ref{subsec:general}). 
For the ease of comparison, we introduce notation 
\mbox{$\Delta (\bf{n_r} \cdot \bm{\sigma}) \equiv J ( \bf{M_r} \cdot \bm{\sigma})$}, 
with unit vector $\bf{n_r}$ pointing in the direction of local magnetization $\bf{M_r}$. 

\begin{figure}
    \centering
    \includegraphics[width = \linewidth]{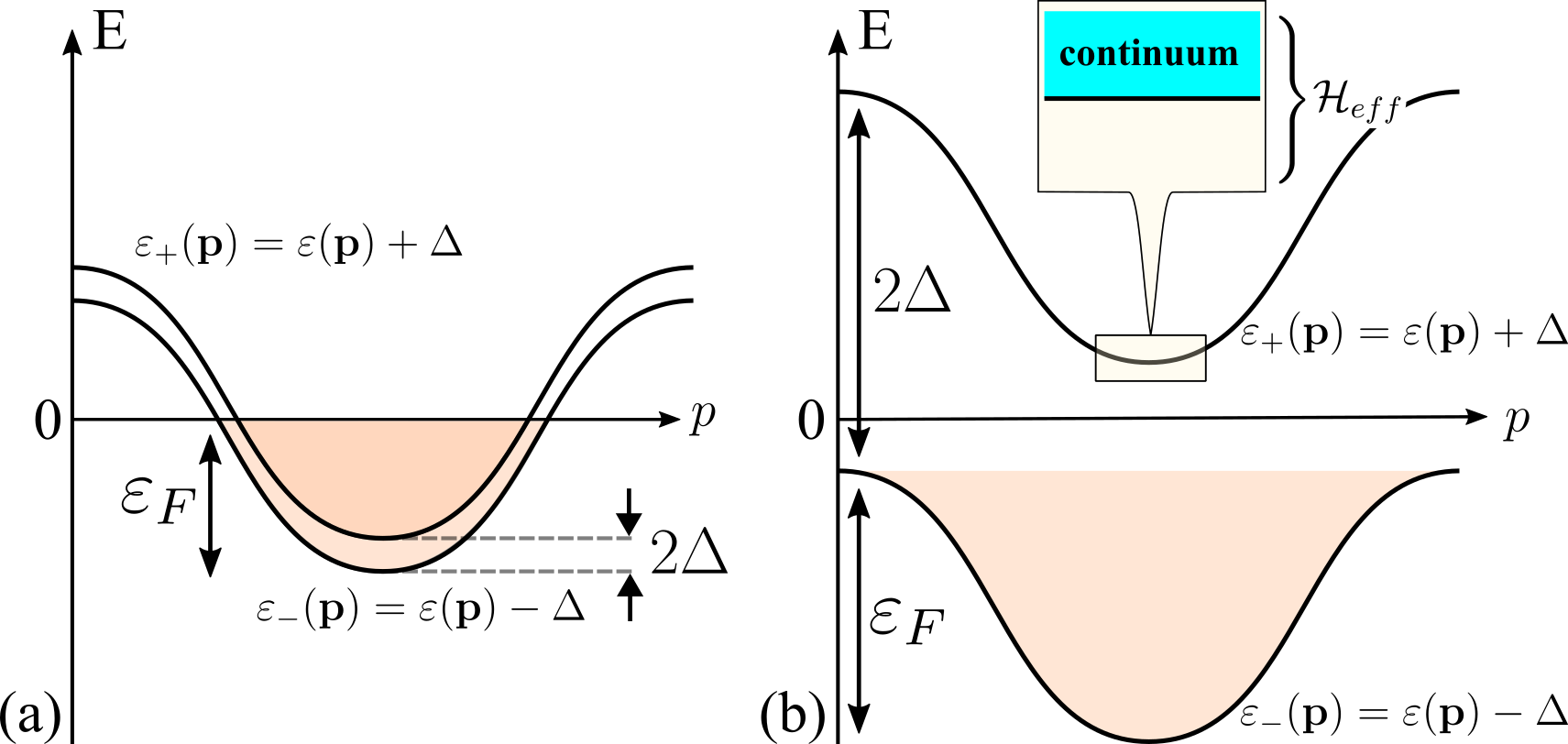}
    \caption{Band dispersion in the uniform ferromagnetic state, where 
    a spin-degenerate band of the parent paramagnetic phase is Stoner-split 
    into two non-degenerate subbands. Panel (a) shows the limit $\Delta \ll \varepsilon_F$, 
    where the minima of both subbands lie deep at the bottom of the Fermi sea, and 
    thus are inaccessible to low-energy and low-temperature experimental probes. 
    Panel (b) shows the limit $2 \Delta \gtrsim \varepsilon_F$ at half-filling, 
    where the lower subband is completely filled while the upper subband is empty. 
    Lightly doping such a state renders the physics of electron states near the 
    upper band minimum accessible to low-energy measurements. Such states can be 
    described by the effective Hamiltonian $\mathcal{H}_{eff}$ of Eq.~\ref{H_eff_FM}.}
    \label{fig: ferromagnetic band dispersion }
\end{figure}

In the uniform state, we choose ${\bf n_r} = {\bf n}_0$ to point along the $z$ axis. 
The band electron spectrum is thus simply
$$
\mathcal{H} = \varepsilon ({\bf p}) + \Delta \sigma_z ,
$$
where $\varepsilon ({\bf p})$ is the notional band dispersion in the absence of ferromagnetism, 
and the second term above describes a uniform Stoner splitting 
\cite{White2007} of $2 \Delta$ between the spin-up and spin-down 
subbands with energies 
$\varepsilon_\pm ({\bf p}) = \varepsilon ({\bf p}) \pm \Delta$, 
see Fig. \ref{fig: ferromagnetic band dispersion }.  

In the presence of a smooth non-uniform texture $\bf{n_r}$, it is convenient 
to employ the unitary transformation of Eq. (\ref{def U}) and recast the Hamiltonian as 
\begin{equation}
  \mathcal{H} = \varepsilon \left( \hat{\bf p} + \mathbf{A}\cdot\bm{\sigma} \right) + \Delta \sigma_z ,
  \label{eq:FM with texture}
\end{equation}
see Eqs. (\ref{Peierls}) and (\ref{H_non_uniforme}). As a result, 
the Stoner splitting term is rendered uniform, and now the texture influences the electron motion via the gauge potential $\mathbf{A}\cdot\bm{\sigma}$. 

A smooth texture induces a perturbation on an energy scale that is small against both $\Delta$ and the Fermi energy $\varepsilon_F$. 
Since we are interested in possible appearance of bound states, let us notice that such states become visible to low-temperature 
measurements only for $\Delta \gtrsim \varepsilon_F$ -- see Fig. \ref{fig: ferromagnetic band dispersion }. 
It is thus fitting to compare our results for shallow bound states in an antiferromagnet with a toy model of a ferromagnet 
with an entirely filled lower subband of width $\varepsilon_F \lesssim 2 \Delta$, 
as shown in Fig. \ref{fig: ferromagnetic band dispersion }(b): at low temperatures, 
a single dopant electron will find itself at low-lying levels of the otherwise empty upper subband, 
which will correspond to the setting that we studied above for an antiferromagnet.

As in Section \ref{subsec:non-uniform}, we are now in a position to write down 
an effective-mass Hamiltonian that will describe low-lying electron states near 
the bottom of the upper subband in 
Fig. \ref{fig: ferromagnetic band dispersion }(b). To this end, notice 
that spin-$x$ and -$y$ components of the vector potential $\mathbf{A}\cdot\bm{\sigma}$ 
produce matrix elements only between the spin-up and spin-down subbands, 
but not \textit{within} either of the two. 
As a result, they can contribute to the effective-mass Hamiltonian 
only via perturbation theory, producing terms of the order of 
$\frac{\hbar^2}{m L^2 \Delta} \ll 1$ relative to those of the first term in Eq. 
 (\ref{eq:FM with texture}). Therefore, to zeroth order in $\hbar^2 / ( m L^2 \Delta )$, the 
 upper-subband ($\sigma_z = +1$) effective-mass Hamiltonian reads 
 \begin{equation}
  \mathcal{H}_\text{eff} = \frac{1}{2m} \left( \hat{p}_i + A_i^z \right)^2 ,
  \label{H_eff_FM}
\end{equation}
where, without loss of generality, we assume isotropic effective mass 
and set zero energy at the bottom of the upper subband. 

Hamiltonian (\ref{H_eff_FM}) is precisely what one would obtain in an antiferromagnet 
upon omitting all but the first term in Hamiltonians (\ref{H_01}) and (\ref{H_0}), 
and keeping to the spin-up problem. Now we will show that, for a $Q = 1$ Belavin-Polyakov 
skyrmion, this Hamiltonian does \textit{not} produce skyrmion-electron bound states, 
contrary to Hamiltonians (\ref{H_01}) and (\ref{H_0}). To this end, consider the problem 
for a Belavin-Polyakov skyrmion in the gauge of Eq. (\ref{eq:m}), that gives rise to $A_i^z$ 
of Eq. (\ref{eq:Az}). Given isotropic mass above, it is convenient to switch to polar 
coordinates $r = \sqrt{x^2 + y^2}$ and $\alpha = \arctan \frac{y}{x}$ as in Subsection 
\ref{subsec: Profile properties}. We also switch to polar components of $A_i^z$, that 
is from $(A_x^z , A_y^z)$ to $(A_r^z , A_\alpha^z)$, 
where $A_r^z = ( \textbf{e}_r \cdot \textbf{A}^z )$ and 
$A_\alpha^z = ( \textbf{e}_\alpha \cdot \textbf{A}^z )$, 
while $\textbf{e}_r = (\cos \alpha , \sin \alpha)$ and 
$\textbf{e}_\alpha = (-\sin \alpha , \cos \alpha)$. 
Using Eq. (\ref{eq:Az}), we find  
\begin{equation}
    A_r^z = 0 , \,\, A_\alpha^z = \frac{\hbar r}{R^2 + r^2}
\end{equation}
Hamiltonian (\ref{H_eff_FM}) now reads
\begin{eqnarray}
  \mathcal{H}_\text{eff}
  \label{eq:fmBP-1}
  & = &
    \frac{\hbar^2}{2 m} 
    \left[ 
    - \frac{1}{r} \frac{\partial}{\partial r} r \frac{\partial}{\partial r}
    + \left( 
    \frac{l}{r} + A_\alpha^z
    \right)^2
    \right] = \\
      \label{eq:fmBP-2}
    & = &
    \frac{\hbar^2}{2 m} 
    \left[ 
    - \frac{1}{r} \frac{\partial}{\partial r} r \frac{\partial}{\partial r}
    + \left( 
    \frac{l}{r} + \frac{r}{R^2 + r^2}
    \right)^2
    \right]
\end{eqnarray}
Here, we took into account $\mathcal{H}_\text{eff}$ being 
independent of $\alpha$, and replaced the orbital momentum operator 
$\hat{l} = - i \frac{\partial}{\partial \alpha}$ by its integer eigenvalue $l = 0, \pm 1, \pm2, ..$. 
We see that $A_\alpha^z$ only modifies the centrifugal energy term 
while keeping it positive-definite. The first term is the kinetic 
energy of a free particle in an $l=0$ state. Hence no bound states, certainly not for 
the $Q=1$ skyrmion above: it is indeed the spin-orbit coupling term $v {\bf A}_y^\| \cdot \bm{\sigma}$ 
in Hamiltonians (\ref{H_01}) and (\ref{H_0}) that generates skyrmion-electron bound states 
in a N\'eel antiferromagnet. 


\section{Gauge transformation}
\label{Appendix : Gauge transformation}

Operator $W$ can be written as 
\begin{equation}
	W = e^{i\phi \mathbf{n} \cdot \bm{\sigma}} = \cos\phi + i\sin\phi\, \mathbf{n}\cdot \bm{\sigma}, 
\end{equation}
leading to  
\begin{equation}
	W^\dagger \partial _i W =  i\partial_i\phi\, \mathbf{n}\cdot \bm{\sigma}  + i\sin\phi \, W^\dagger \partial_i  \mathbf{n}\cdot \bm{\sigma}
\end{equation}
and then to 
\begin{equation}
	\begin{split}
		&\tilde{\mathbf{A}_i} = -i\hbar U^\dagger W^\dagger \partial _i \left( WU \right) = \mathbf{A}_i + \hbar \partial_i\phi \sigma^z\\ 
		&+ \hbar U^\dagger \left[ \frac{\sin2\phi}{2} \partial_i \mathbf{n}\cdot \bm{\sigma} + \sin^2\phi (\mathbf{n} \times \partial_i \mathbf{n} )\cdot \bm{\sigma} \right]U. \label{2nd gauge transformation}
	\end{split}
\end{equation}
The last term lies in the plane orthogonal to $\mathbf{e}_z$, and thus only contributes to the transformation of $\mathbf{A}_i^{\|}$. 
 
Acting with $\partial_i$ on Eq. (\ref{def U}) yields
\begin{equation}
		\partial_iU^\dagger \mathbf{n}\cdot \bm{\sigma} U + U^\dagger \partial_i\mathbf{n}\cdot \bm{\sigma} U + U^\dagger \mathbf{n}\cdot \bm{\sigma} \partial_iU = 0. 
\end{equation}
A right multiplication by $U^\dagger \mathbf{n}\cdot \bm{\sigma} U = \sigma^z$ then leads to 
\begin{equation}
	\partial_iU^\dagger U - i U^\dagger(\mathbf{n}\times \partial_i \mathbf{n})\cdot \bm{\sigma} U + \sigma^z U^\dagger\partial_iU \sigma^z = 0.
\end{equation}
Finally, the identity $\partial_i\left(U^\dagger U\right) = 0$ allows to write 
\begin{equation}
	U^\dagger\partial_i U - \sigma^z U^\dagger\partial_iU \sigma^z = - i U^\dagger(\mathbf{n}\times \partial_i \mathbf{n})\cdot \bm{\sigma} U .
\end{equation}
Which, by definition of gauge field $A_i^\alpha$ in Eq. (\ref{Peierls}), means that its in-plane components reads  
\begin{equation}
	\mathbf{A}_i^\parallel = - \frac{\hbar}{2} U^\dagger\left[(\mathbf{n}\times \partial_i \mathbf{n})\cdot \bm{\sigma} \right] U.
\end{equation}

Together with (\ref{2nd gauge transformation}), this implies 
\begin{equation}
	\begin{split}
		\tilde{\mathbf{A}}_i^{\|} = \hbar U^\dagger \left[\frac{\sin2\phi}{2} \partial_i \mathbf{n}\cdot \bm{\sigma} \right. \hspace{3cm} \\
		+ \left. \left(\sin^2\phi - \frac{1}{2} \right) (\mathbf{n} \times \partial_i \mathbf{n} )\cdot \bm{\sigma}\right]U .
	\end{split} 
\end{equation}
Using the identity $\sin^2 \phi - \frac{1}{2} = - \frac{1}{2} \cos 2 \phi$, we then find 
\begin{equation}
		\left(\tilde{\mathbf{A}}_i^{\|}\right)^2 = \frac{\hbar^2}{4} ||\partial_i\mathbf{n}||^2 = \left(\mathbf{A}_i^{\|}\right)^2.
\end{equation}
The norm $\left(\mathbf{A}_i^{\|}\right)^2$ being conserved, the in plane components $\mathbf{A}_i^{\|}$ thus only undergo an in-plane rotation, as they do under gauge transformation by operator $V_z$. \\

From the gauge properties above, we can deduce the allowed terms 
\begin{eqnarray}
		\left(\mathbf{A}_i^\parallel\right)^n &\to& \left(\tilde{\mathbf{A}}_i^\parallel\right)^n \nonumber \\
		\left(\hat{p}_i + A_i^z\sigma^z\right)^n &\to& \left(\hat{p}_i + \tilde{A}_i^z\sigma^z\right)^n\\
		\{\hat{p}_i,\mathbf{A}_i^\parallel \cdot\bm{\sigma} \} &\to& \{\hat{p}_i + \delta A_i^z\sigma^z,\tilde{\mathbf{A}}_i^\parallel \cdot\bm{\sigma} \} = \{\hat{p}_i,\tilde{\mathbf{A}}_i^\parallel \cdot\bm{\sigma} \} \nonumber
\end{eqnarray}
where in the last line we used the identity $\{\sigma^i,\sigma^j \} = 0$ for $i\neq j$. We observe that all the terms present in $\hat{\bm{\gamma}}_\pm$ in Eqs. (\ref{gammas}) are thus allowed, meaning that Hamiltonian (\ref{H complete 4x4}) and its low-energy limit (\ref{H_0}) are indeed gauge-invariant.

\section{Gauge invariants}
\label{Appendix : Gauge invariants}
In this Appendix, we set $\hbar = 1$. 
In Subsection \ref{subsec:transformations}, we saw that gauge transformations (\ref{transformation de jauge}) and (\ref{transformation de jauge 2}) amount only to local rotation of in-plane components $\mathbf{A}_i^\parallel$, thus conserving 
$\left( \mathbf{A}_i^\parallel \right)^2$ as well as $\left[ \mathbf{A}_x^\parallel \times \mathbf{A}_y^\parallel \right]$ 
and $\left( \mathbf{A}_x^\parallel \cdot \mathbf{A}_y^\parallel \right)$.
The absolute value $\left(\mathbf{A}^\parallel\right)^2 = \sum_{i=x,y}\left( \mathbf{A}_i^\parallel \right)^2$ can be expressed as (with implicit sum over repeated indices)
\begin{equation}
	4\left(\mathbf{A}^\parallel\right)^2 = \left(\partial_i\theta \partial^i\theta + \sin^2\theta\, \partial_i\phi\partial^i\phi\right) = \left(\bm{\nabla}\cdot \mathbf{n}\right)^2,
\end{equation}
that is the gradient energy density of the isotropic antiferromagnet. 
Vector product $\left[ \mathbf{A}_x^\parallel \times \mathbf{A}_y^\parallel \right]$ can be written as 
\begin{equation}
	\begin{split}
		\mathbf{A}_x^\parallel \times \mathbf{A}_y^\parallel &= \frac{1}{4} \sin\theta \left(\partial_x\theta \partial_y\phi - \partial_y\theta \partial_x \phi\right)\\ 
		&= \frac{1}{4} \left( \mathbf{n} \cdot \left[ \partial_x\mathbf{n} \times \partial_y \mathbf{n} \right] \right)
  = \pi \mathcal{D}_s(\mathbf{r}),
	\end{split}
\end{equation}
where $\mathcal{D}_s$ is the skyrmion density \cite{Tatara_2008,Jiang2017}, defined so that
\begin{equation}
	\int d^2r \mathcal{D}_s(\mathbf{r}) = \int d^2r \frac{1}{4\pi}\left(\partial_x\mathbf{n} \times \partial_y \mathbf{n}\right) \cdot \mathbf{n} = \mathcal{N}_s
\end{equation}
where $\mathcal{N}_s \in \mathbb{Z}$ is the total topological charge of skyrmions present in the plane. 
Another invariant, suggested by the transformation law of $A_i^z$ in Eq. (\ref{transformation de jauge}) 
is the fully antisymmetric tensor 
\begin{equation}
	F^z_{ij} = \partial_iA_j^z - \partial_jA_i^z.
\end{equation}
The latter invariant can be recast as
\begin{equation}
	\begin{split}
		F_{ij}^z &= \partial_i\left(\partial_j\phi \frac{1-\cos\theta}{2}\right) - \partial_j\left(\partial_i\phi \frac{1-\cos\theta}{2}\right) \\
		&= \frac{1}{2}\sin\theta \left(\partial_i\theta\partial_j\phi - \partial_j\theta\partial_i\phi\right) 
                =  2\pi \epsilon_{ij} \mathcal{D}_s(\mathbf{r}) ,\\
		\label{invariant de jauge}
	\end{split}
\end{equation} 
with $\epsilon_{ij}$ the fully antisymmetric tensor. This means that there are only two distinct invariants associated with gauge transformation (\ref{transformation de jauge}).  Note that it also implies that the textbook non-abelian gauge field tensor is identically equal to zero: 
\begin{equation}
	\mathbf{F}_{ij} \equiv \partial_i \mathbf{A}_j - \partial_j\mathbf{A}_i + i\left[\mathbf{A}_i,\mathbf{A}_j\right] =0.
\end{equation}
This is due to the "electromagnetic" part exactly cancelling the anti-commutator, both being proportional to skyrmion density.

\section{Isotropic toy model}
\label{Appendix: Isotropic toy model}

In this appendix, we show that the loss of bound states at small BP skyrmion radia is not related 
to the mass anisotropy in Eq. (\ref{eq:anisotropy}). A perfectly isotropic toy model allows us to 
capture the scale $\bar{R} = \sqrt{\epsilon}\xi$, below which the bound states are lost. To this 
end, consider Hamiltonian (\ref{H_0}) with isotropic mass 
\begin{equation}
	H_i =  \sum_{i=x,y}\frac{1}{2m^*}\left(\hat{p}_i + A_i^z\sigma^z\right)^2 + v  A_{y}^x\sigma_x. \label{H isotrope}
\end{equation}
The repulsive potential 
\begin{equation}
	\frac{1}{2m^*}\sum_i \left(A_i^z\right)^2 = \frac{\hbar^2}{2m^*}\frac{(1-\cos\theta)^2}{4r^2}
\end{equation} 
does not depend on polar angle $\alpha$, while the cross term is proportional to
\begin{equation}
	 \sum_iA_i^z \partial_i = \frac{(1-\cos \theta) }{2r} \mathbf{e}_\alpha \cdot \bm{\nabla } = \frac{(1-\cos \theta) }{2r^2}  \frac{\partial}{\partial \alpha}
\end{equation}
and thus only acts on $\alpha$. That is, Hamiltonian (\ref{H isotrope}) allows separation of variables, 
and the angular momentum $l$ is a good quantum number. For $l=0$, Hamiltonian (\ref{H isotrope}) reads
\begin{equation}
	\begin{split}
		H_i^{l=0} =&\; \frac{\hbar^2}{2m^*R^2} \left( -\frac{1}{z} \frac{\partial}{\partial z}\left(z \frac{\partial}{\partial z}\right) + \frac{z^2}{(1+z^2)^2} \right) \\ 
        &- \frac{\hbar v}{R} \frac{1}{1+z^2}\sigma^x.
        \label{H isotrope l=0}
	\end{split}
\end{equation} 
To keep the same density of states as for Hamiltonian (\ref{H_0}), we choose the effective mass $m^* = \sqrt{m_x^*m_y^*}$. 
Now treat Hamiltonian (\ref{H isotrope l=0}), considering the potential as a perturbation relative to the kinetic energy \cite{Landau1981Quantum}. To this end, we find the wave function in two separate regions. First, denote the sum of the 
potential terms in Eq.~(\ref{H isotrope l=0}) as $U(z)$, and consider the skyrmion core, where $|\varepsilon| \ll |U(z)|$: 
the Schrödinger equation reads
\begin{equation}
	\frac{\hbar^2}{2m^*R^2} \frac{1}{z} \frac{d}{dz} \left(z\frac{d\psi}{dz}\right) = U(z) \psi(z).
\end{equation}
The above equation can now be integrated along $z$ to the upper limit $z_1$ of the attractive region: 
\begin{equation}
	\left. \frac{d\psi}{dz} \right|_{z_1} \simeq \frac{1}{z_1} \frac{2m^*R^2}{\hbar^2} \psi(z_1)\int_0^{z_1} U(z)zdz. 
 \label{eq : Iso wave function matching}
 \end{equation}
Here we took advantage of the observation that the sought wave function is spread out, and does not vary much within the potential well. By contrast, far from the attractive region, i.e. for $z \gg 1$, the Schr\"odinger equation describes a free particle, whose wave function is proportional to the Hankel function $H_0^{(1)}(i\kappa z)$, with $\kappa = \sqrt{\frac{2m^*R^2|\varepsilon|}{\hbar^2}} \ll 1$. 
For $\kappa z \ll 1$, this function behaves asymptotically as $\ln(\kappa z)$. The value of $z_1$ being of the order of unity 
and thus $\kappa z_1$ being small allows us to match the logarithmic derivative $\frac{1}{\psi} \frac{d \psi}{dz}$ of this asymptotic form to that in Eq. (\ref{eq : Iso wave function matching}) at $z = z_1$, which yields  
\begin{equation}
	\begin{split}
		\varepsilon  &\simeq - \frac{\hbar^2}{2m^*R^2z_1^2}\exp\left[\frac{\hbar^2}{m^*R^2}\left(\int_0^{z_1} U(z)zdz\right)^{-1}\right] \\
		&\simeq - \Delta\frac{R_0}{R}\frac{\xi}{R}\frac{e^{2/\ln z_1}}{z_1^2}\exp\left[\frac{1}{1-\frac{R}{R_0}}\right]
	\end{split}
\end{equation}
with $R_0 \equiv \frac{\sqrt{\epsilon}}{2}\xi$. The expression above demonstrates the existence of a skyrmion radius $\bar{R} \sim R_0$, where the lowest bound state merges into the continuum and disappears. The same result can be obtained by solving Hamiltonian (\ref{H isotrope l=0})  numerically.

\bibliographystyle{IEEEtran}
\bibliography{IEEEabrv,Biblio_these}

\end{document}